\documentclass[reprint,aps,prd,floatfix,nofootinbib,showkeys,showpacs,superscriptaddress]{revtex4-2}
\usepackage{amsmath,amsthm,amssymb, mathrsfs}
\usepackage{graphicx,color,dsfont,bm, url, hyperref}
\usepackage[utf8]{inputenc}
\usepackage{multirow}
\usepackage{booktabs}
\usepackage{siunitx}
\newcommand{\asb}{\bar{\alpha}_s}
\def\X{{\scriptscriptstyle X}}
\def\R{{\scriptscriptstyle\mathrm{R}}}
\def\V{{\scriptscriptstyle\mathrm{V}}}
\def\MC{{\scriptscriptstyle\mathrm{MC}}}
\def\inn{\text{in}}
\def\out{\text{out}}

\def\CF{\mathrm{C_F}}
\def\CFsq{\mathrm{C_F^2}}
\def\CFcub{\mathrm{C_F^3}}

\def\CA{\mathrm{C_A}}
\def\CAsq{\mathrm{C_A^2}}
\def\CAcub{\mathrm{C_A^3}}

\def\cA{\mathcal{A}} \def\cAb{\mathcal{\bar{A}}}
\def\cB{\mathcal{B}} 
\def\cC{\mathcal{C}}

\def\cF{\mathcal{F}}
\def\cG{\mathcal{G}}

\def\cO{\mathcal{O}}
\def\cQ{\mathcal{Q}}
\def\cS{\mathcal{S}}

\def\cW{\mathcal{W}}

\def\O{\Omega}
\def\Ob{\bar{\Omega}}

\def\Uh{\hat{\mathcal{U}}}

\def\aktt{\scriptscriptstyle\mathrm{ak_t}}
\def\ktt{\scriptscriptstyle\mathrm{k_t}}
\def\caa{\scriptscriptstyle\mathrm{C/A}}
\begin{document}

\title{$V/H$+Jet Production with the Cambridge/Aachen Algorithm}

\author{K. \surname{Khelifa-Kerfa}}
\affiliation{Department of Physics, Faculty of Science and Technology, Relizane University, Relizane 48000, Algeria}
\affiliation{Laboratory of Physics of Experimental Techniques and Applications, University of Medea, Medea 26000, Algeria}
\affiliation{Laboratory of Thin Layers and Advanced Technologies, Relizane University, Relizane 48000, Algeria}
\affiliation{Laboratory of Mathematics and Applications, University of Chlef, Chlef 02000, Algeria}

\email{kamel.khelifakerfa@univ-relizane.dz}

\begin{abstract}
We present fixed-order perturbative calculations up to four-loop order for a generic non-global QCD observable in hadron-hadron collisions. Specifically, we study the invariant-mass distribution of the  highest-$p_t$ jet produced in association with a vector boson or a  Higgs boson, where jets are defined using the Cambridge/Aachen sequential recombination algorithm. This work is part of a series of  papers~\cite{Khelifa-Kerfa:2015mma, Khelifa-Kerfa:2024hwx,  Khelifa-Kerfa:2024dut, Khelifa-Kerfa:2025cdn, Khelifa-Kerfa:2024udm,  Khelifa-Kerfa:2025jev} examining the impact of various jet algorithms  on the perturbative structure of non-global observables across different collision environments. We find that at fixed order the Cambridge/Aachen  algorithm suppresses the large non-global logarithms more effectively  than the anti-$k_t$ and $k_t$ algorithms, while at all orders it  performs comparably to the $k_t$ algorithm. Furthermore, comparisons of all-orders resummed form factors reveal that finite-$N_c$ corrections remain at the percent level, consistent with previous findings.
\end{abstract}

%\pacs{12.38.Bx,11.80.Fv,13.66.Bc}

\maketitle

%%%%%%%%%%%%%%%%%%%%%%%%%%%%%%%%%%%%%%%%%%%%%%%%%%%%%%%%%%%%%
\section{Introduction}
\label{sec:Intro}

Precision calculations of jet observables at hadron colliders, particularly at the Large Hadron Collider (LHC) at CERN, demand an
increasingly refined understanding of QCD radiation patterns. Among the most phenomenologically important of these observables is the invariant mass of a jet produced in association with an electroweak boson or a Higgs boson --- a configuration that serves simultaneously as a stringent test of the Standard Model (SM) and as a sensitive probe of new physics through its jet substructure~\cite{Kogler:2018hem, Larkoski:2017jix}. The invariant jet-mass distribution is especially valuable because it encodes information about color flow, underlying-event activity, hadronization, and the interplay of multiple logarithmic hierarchies---all of which are directly relevant to precision measurements at the LHC.

In the limit of small normalized jet mass, $m_j^2/p_{t}^2 \ll 1$, fixed-order perturbation theory becomes inadequate owing to
the proliferation of logarithmically enhanced terms. These arise from multiple sources. The first category, commonly termed \emph{global logarithms}, comprises contributions from soft-collinear and hard-collinear radiation off all incoming and outgoing hard partons; upon resummation these exponentiate into the well-known Sudakov form factor. The second, phenomenologically more challenging, category comprises \emph{non-global logarithms} (NGLs)~\cite{Dasgupta:2001sh, Dasgupta:2002bw}, which arise from soft-gluon correlations connecting spatially separated phase-space regions --- specifically, secondary emissions from inside the jet that are induced by harder radiation deposited outside. NGLs are a manifestation of the genuinely non-Abelian color dynamics of QCD and resist straightforward exponentiation. Despite sustained theoretical efforts~\cite{Banfi:2002hw, Schwartz:2014wha, Khelifa-Kerfa:2015mma, Hatta:2013iba, Banfi:2021owj, Banfi:2021xzn, Becher:2023vrh, Becher:2023mtx}, their complete resummation remains an open problem, particularly beyond the leading-color approximation.

A third class of large logarithms, known as \emph{clustering logarithms} (CLs), arises whenever jets are defined by sequential recombination algorithms other than the anti-$k_t$ one~\cite{Cacciari:2008gp}. First recognized in the context of the $k_t$ algorithm by Banfi and Dasgupta~\cite{Banfi:2005gj}, and investigated further in Refs.~\cite{Appleby:2002ke,Delenda:2006nf}, these contributions stem from the algorithm-induced recombination of Abelian, independent primary gluon emissions. Such recombination distorts the real-virtual cancellations that ordinarily occur for primary emissions, thereby introducing a new tower of single logarithms at each perturbative order. Crucially, whereas clustering suppresses the magnitude of the standard NGLs --- by preventing certain gluon configurations from contributing to the jet mass --- it simultaneously generates CLs~\cite{Banfi:2010pa,Khelifa-Kerfa:2011quw,Delenda:2012mm}, so that the total logarithmic content of the observable cannot be reduced simply by a judicious choice of algorithm.

The Cambridge/Aachen (C/A) algorithm~\cite{Dokshitzer:1997in,Wobisch:1998wt} occupies a distinguished position among sequential recombination schemes. Unlike the $k_t$ algorithm~\cite{Catani:1993hr,Ellis:1993tq} ---whose distance measure weights angular separations by the smaller of the two particles' transverse momenta, causing the softest emissions to merge first --- the C/A algorithm employs a purely geometric distance measure that is independent of any energy scale. This angular-ordered clustering sequence closely mirrors the structure of QCD coherence~\cite{Salam:2010nqg} and is widely exploited in jet substructure studies, particularly in the context of jet grooming~\cite{Larkoski:2017jix}.
It is also the structural foundation of modern jet-grooming techniques: since the C/A clustering history is purely angular-ordered, it can be traversed backward from wide angle to narrow angle, providing the well-defined declustering sequence on which groomers such as the modified mass-drop tagger and soft drop operate~\cite{Larkoski:2017jix,Dasgupta:2013ihk,Larkoski:2014wba}.
Furthermore, among the three members of the generalized sequential recombination family --- anti-$k_t$~\cite{Cacciari:2008gp}, $k_t$, and C/A --- the C/A algorithm has been shown to yield the smallest non-global logarithmic contributions~\cite{Banfi:2010pa,Khelifa-Kerfa:2011quw,Khelifa-Kerfa:2025jzl}, making it theoretically attractive for analyses where logarithmic accuracy is paramount.

Analytic understanding of NGLs and CLs in hadronic processes has advanced significantly in recent years. Fixed-order calculations of NGLs for the jet-mass distribution in $Z$+jet and dijet production at two loops were presented in~\cite{Dasgupta:2012hg}, while the companion paper~\cite{Ziani:2021dxr} extended these to include both NGLs and CLs in the $k_t$ and C/A algorithms, also at two loops, for Higgs- and vector-boson associated jet production. Multi-loop analyses were subsequently carried out for $e^+e^-$ annihilation processes~\cite{Khelifa-Kerfa:2015mma,Schwartz:2014wha,Delenda:2012mm,Khelifa-Kerfa:2024hwx,Khelifa-Kerfa:2024dut,Khelifa-Kerfa:2024roc}, revealing that the fixed-order CL series exponentiates and that the perturbative expansion converges rapidly. For hadronic collisions, a systematic study of NGLs up to four loops with full color dependence and complete jet-radius dependence in the anti-$k_t$ algorithm was reported in~\cite{Khelifa-Kerfa:2024udm} for $V/H$+jet processes, while the companion work~\cite{Khelifa-Kerfa:2025jev} extended these calculations to the $k_t$ algorithm, providing the first four-loop results encompassing both CLs and NGLs for the jet mass in a hadronic environment.

The present paper continues this program by extending the work of~\cite{Khelifa-Kerfa:2025jev} to the Cambridge/Aachen algorithm. We compute the complete fixed-order distribution of the normalized invariant mass squared of the leading jet in $V/H$+jet production, incorporating contributions from both CLs and NGLs through four-loop order in perturbation theory. Our calculations are carried out in the eikonal approximation with strong energy ordering among the final-state soft gluons, ensuring single-logarithmic accuracy. By retaining the full dependence on color factors and the jet-radius parameter $R$ for all relevant partonic channels, we provide semi-analytical expressions that are directly applicable to phenomenological studies.

Several notable features distinguish the C/A case from both the anti-$k_t$ and $k_t$ results. First, as is also the case at two loops, the C/A algorithm yields NGL and CL coefficients that are identical to those of $k_t$ clustering. Second, and in contrast to the $k_t$ situation studied in~\cite{Khelifa-Kerfa:2025jev}, the purely geometric nature of the C/A distance measure introduces distinct higher-loop clustering structures starting at three-loop order, which we derive and analyze in detail. The resulting NGL and CL contributions are smaller in magnitude than in the $k_t$ case, and smaller still compared to the anti-$k_t$ case. Third, we find that the NGL and CL coefficients exhibit the universal ``boundary'' or ``edge'' effect --- approaching finite non-zero constants as $R \to 0$ --- consistent with observations in $e^+e^-$ calculations~\cite{Khelifa-Kerfa:2011quw,Delenda:2012mm,Khelifa-Kerfa:2024hwx} and in earlier hadronic results~\cite{Ziani:2021dxr,Khelifa-Kerfa:2024udm,Khelifa-Kerfa:2025jev}. Fourth, the all-orders analysis of C/A clustering shows that it produces non-global effects that are almost identical in size to the $k_t$ case, particularly for values of the jet-mass observable where the exponentiation of the fixed-order results is reliable. This observation is further confirmed by a detailed study of the specific $Z$+jet process. These features make the C/A and $k_t$ algorithms preferable to anti-$k_t$ for mitigating the impact of non-global logarithms. Fifth, the finite-$N_c$ corrections are shown to remain at the percent level for phenomenologically relevant values of the observable. This is in excellent agreement with earlier results reported for both lepton and hadron processes (see, for instance,~\cite{Hatta:2013iba, Hatta:2020wre, Khelifa-Kerfa:2024udm}).

It is worth noting that, unlike the cases of the anti-$k_t$ and $k_t$ algorithms, the all-orders numerical resummation of non-global observables defined with the C/A algorithm is severely limited by the availability of dedicated codes. The only available code for C/A clustering is that of Ref.~\cite{Becher:2023znt}, which, however, is restricted to observables with purely soft singularities, such as the energy-flow (or gap-between-jets) observable. The invariant jet-mass observable exhibits both soft and collinear singularities and therefore falls outside the scope of that code. The Dasgupta--Salam Monte Carlo (MC) code~\cite{Dasgupta:2001sh}, upon which the code of Ref.~\cite{Becher:2023znt} is based, covers observables with both soft and collinear singularities but is restricted to the anti-$k_t$ and $k_t$ algorithms, with no implementation for C/A. Consequently, no all-orders numerical results are available against which to benchmark our analytical C/A calculations; we therefore compare instead to the $k_t$ results produced by the MC code of~\cite{Dasgupta:2001sh}.

Throughout the calculation we treat super-leading logarithms (SLLs)~\cite{Forshaw:2006fk,Becher:2021zkk,Becher:2023mtx} as decoupled from the standard NGL and CL contributions at the loop orders considered here; any potential interleaving between these two classes first enters at five-loop order for $2 \to 1$ processes~\cite{Becher:2021zkk,Becher:2023mtx,Khelifa-Kerfa:2024udm}.

The paper is organized as follows. Section~\ref{sec:Defs} introduces the kinematic framework, the observable definition, the C/A jet algorithm, and the structure of the resummed jet-mass distribution. Section~\ref{sec:FO} presents the three- and four-loop calculations of CLs and NGLs for all partonic channels. Section~\ref{sec:AllOrders} compares the fixed-order analytical predictions against all-orders numerical resummation from the Monte Carlo code of~\cite{Dasgupta:2001sh}. Section~\ref{sec:Conclusion} summarizes our conclusions and discusses prospects for further work.
%%%%%%%%%%%%%%%%%%%%%%%%%%%%%%%%%%%%%%%%%%%%%%%%%%%%%%%%%%
\section{Definitions}
\label{sec:Defs}

\subsection{Kinematics}
\label{subsec:Kinematics}

We study the production of a single hard jet in association with a Higgs boson $H$ or a vector boson $V \in \{Z, W^\pm, \gamma\}$ at a proton--proton collider. The relevant partonic Born channels are identical to those discussed in~\cite{Khelifa-Kerfa:2024udm,Khelifa-Kerfa:2025jev}: for vector-boson-associated production there are two topologically distinct processes,
\begin{subequations}
\begin{equation}
(\delta_1):\; q\bar{q} \to g\,V/H, \qquad (\delta_2):\; qg \to q\,V/H,
\label{eq:Born12}
\end{equation}
while Higgs production acquires an additional all-gluon channel,
\begin{equation}
(\delta_3):\; gg \to g\,H.
\label{eq:Born3}
\end{equation}
\end{subequations}
The channel $\bar{q}g \to \bar{q}\,V/H$ is color-equivalent to $(\delta_2)$ and need not be tracked separately. From the perspective of our soft-gluon calculations, all three channels involve exactly three hard colored legs together with a color-singlet boson; they differ only through their Born cross sections and the corresponding color Casimir scalars. The Higgs-boson channel $gg \to gH$ proceeds via an effective vertex encoding the heavy top-quark loop.

We denote the four-momenta of the incoming partons by $p_a$ and $p_b$, the outgoing hard parton by $p_j$, and those of the subsequently emitted soft gluons by $k_i$ ($i = 1, \ldots, m$). In the center-of-mass frame these are parametrized as
\begin{subequations}
\begin{equation}
p_a = x_a\frac{\sqrt{s}}{2}(1,0,0,1), \qquad p_b = x_b\frac{\sqrt{s}}{2}(1,0,0,-1),
\label{eq:pa_pb}
\end{equation}
and
\begin{align}
p_j &= p_t(\cosh y,\,\cos\varphi,\,\sin\varphi,\,\sinh y), \label{eq:pj} \\
k_i &= k_{ti}(\cosh\eta_i,\,\cos\phi_i,\,\sin\phi_i,\,\sinh\eta_i),
\label{eq:ki}
\end{align}
\end{subequations}
where $\sqrt{s}$ is the partonic center-of-mass energy; $x_a$ and $x_b$ are the momentum fractions carried by the colliding partons, described by standard parton distribution functions currently available through next-to-next-to-next-to-leading-order (N$^3$LO) accuracy~\cite{Barontini:2024eii}; $p_t$, $y$, and $\varphi$ are the transverse momentum, rapidity, and azimuthal angle of the outgoing hard parton with respect to the beam axis; and $k_{ti}$, $\eta_i$, and $\phi_i$ are the corresponding kinematic quantities for the $i$-th soft gluon. All hard partons are treated as massless; extensions incorporating heavy-quark mass effects are reserved for future work. Recoil corrections enter at next-to-next-to-leading logarithmic accuracy and are accordingly neglected throughout~\cite{Banfi:2004yd}.

To facilitate the angular integrations, it is convenient to introduce polar variables $(r_i, \theta_i)$ centered on the hard-jet axis:
\begin{equation}
\eta_i - y = R\,r_i\cos\theta_i, \qquad \phi_i - \varphi = R\,r_i\sin\theta_i,
\label{eq:polar}
\end{equation}
with $r_i > 0$ and $0 < \theta_i < 2\pi$, so that $\mathrm{d}\eta_i\,\mathrm{d}\phi_i = R^2 r_i\,\mathrm{d}r_i\,\mathrm{d}\theta_i$.

The one-loop dipole antenna function, encoding the universal pattern of soft-gluon radiation from a color dipole $(\alpha,\beta)$, is given by
\begin{equation}
w_{\alpha\beta}^i = \frac{k_{ti}^2}{2}\frac{(p_\alpha \cdot p_\beta)}{(p_\alpha \cdot k_i)(k_i \cdot p_\beta)}.
\label{eq:antenna}
\end{equation}
The corresponding color factors for the three Born dipoles $\Delta_\delta = \{(aj),(bj),(ab)\}$ in channel $\delta$ read~\cite{Khelifa-Kerfa:2020nlc,Delenda:2015tbo}
\begin{equation}
\cC_{q\bar{q}} = \cC_{qq} = 2\CF - \CA, \qquad \cC_{qg} = \cC_{gg} = \CA,
\label{eq:dipolecolor}
\end{equation}
where the color Casimir scalars are $\CF = (N_c^2-1)/(2N_c)$ and $\CA = N_c$, with $N_c = 3$ in QCD.

%========================================
\subsection{Observable and jet algorithm}
\label{subsec:Observable+jetAlgo}

The observable under study is the normalized invariant mass squared of the leading hard jet,
\begin{equation}
\varrho = \frac{m_j^2}{p_t^2} = \frac{1}{p_t^2}\!\left(p_j + \sum_{i\in j} k_i\right)^{\!2} = \sum_{i\in j}\varrho_i + \mathcal{O}\!\left(\frac{k_{ti}^2}{p_t^2}\right),
\label{eq:rho_def}
\end{equation}
where the sum runs over all soft gluons recombined into the jet by the clustering procedure. The individual contribution of gluon $k_i$ to $\varrho$ is
\begin{equation}
\varrho_i = \frac{2(p_j \cdot k_i)}{p_t^2} = 2\xi_i\bigl[\cosh(\eta_i - y) - \cos(\phi_i - \varphi)\bigr],
\label{eq:rho_i}
\end{equation}
where $\xi_i \equiv k_{ti}/p_t$ and terms of order $k_{ti}^2/p_t^2$ have been dropped in the eikonal limit. Expanding in powers of the jet radius $R$ via the parametrization~\eqref{eq:polar} yields
\begin{equation}
\varrho_i = \xi_i\!\left[R^2 r_i^2 + \frac{R^4 r_i^4}{12}\cos(2\theta_i) + \cdots\right],
\label{eq:rho_expand}
\end{equation}
and retaining only the leading piece --- which contributes at single-logarithmic accuracy --- gives
\begin{equation}
\varrho_i \simeq R^2 \xi_i r_i^2.
\label{eq:rho_leading}
\end{equation}

The Cambridge/Aachen (C/A) jet algorithm~\cite{Dokshitzer:1997in,Wobisch:1998wt} is a member of the inclusive generalized sequential recombination family~\cite{Cacciari:2011ma}, parametrized by the inter-particle and beam distances
\begin{equation}
d_{ij} = \frac{\Delta R_{ij}^2}{R^2}, \qquad d_{iB} = 1,
\label{eq:CA_distances}
\end{equation}
where $\Delta R_{ij}^2 = (\eta_i-\eta_j)^2 + (\phi_i-\phi_j)^2$. The algorithm proceeds iteratively: at each step one identifies the smallest distance among all $\{d_{ij}, d_{iB}\}$; if it corresponds to a pair $(i,j)$, the two particles are merged into a pseudo-jet via the $E$-scheme recombination, in which the four-momenta of the two particles are added directly; if it corresponds to a beam distance $d_{iB}$, particle $i$ is promoted to a final-state jet and removed from the list. This cycle continues until no particles remain.

The defining characteristic of the C/A algorithm --- distinguishing it from both the $k_t$~\cite{Catani:1993hr,Ellis:1993tq} and anti-$k_t$~\cite{Cacciari:2008gp} schemes --- is that its distance measure is entirely geometric, carrying no dependence on the particles' transverse momenta. Consequently, the clustering sequence is governed purely by angular proximity: the pair with the smallest angular separation is always merged first, irrespective of the relative softness of the two constituents.

By contrast, the $k_t$ algorithm weights angular separations by $\min(p_{ti}^2, p_{tj}^2)$, ensuring that the softest particles cluster earliest. In the eikonal approximation with strong energy ordering $\xi_1 \gg \xi_2 \gg \cdots \gg \xi_m$, the pairwise condition for two soft gluons to merge before either is individually declared a jet takes the same geometric form for both the $k_t$ and C/A algorithms:
\begin{equation}
\Delta R_{ij}^2 < R^2 \quad\Longleftrightarrow\quad r_i^2 + r_j^2 - 2r_ir_j\cos(\theta_i-\theta_j) < 1.
\label{eq:CA_pair_condition}
\end{equation}
The two algorithms therefore produce identical results at two-loop order, where only a single pair of soft gluons needs to be considered. The genuinely distinct behavior of the C/A scheme emerges at three-loop order and beyond, where the ordering of clustering steps in a multi-particle configuration can differ significantly. Specifically, in the $k_t$ algorithm, the softest emission always clusters first with its nearest neighbor --- the energy-ordered clustering priority --- while in C/A the angularly closest pair clusters first regardless of the energy hierarchy. This distinction generates different irreducible clustering structures at three-loop order and beyond, which are the primary focus of the present work.
%========================================
\subsection{Observable distribution}
\label{subsec:ObservableDist}

At next-to-leading logarithmic (NLL) accuracy, the differential jet-mass cross section in a partonic channel $\delta$ takes the factorized form~\cite{Dasgupta:2012hg,Ziani:2021dxr,Khelifa-Kerfa:2024udm}
\begin{equation}
\frac{\mathrm{d}\Sigma_\delta(\rho)}{\mathrm{d}\mathcal{B}_\delta} = \int_0^\rho \frac{\mathrm{d}^2\sigma_\delta}{\mathrm{d}\mathcal{B}_\delta\,\mathrm{d}\varrho}\,\mathrm{d}\varrho,
\label{eq:dsigma}
\end{equation}
where $\rho$ is the jet-mass veto, $\mathcal{B}_\delta$ encodes the Born phase space including parton distribution functions and the hard-scattering matrix element squared, and $\mathrm{d}\sigma_\delta$ is the fully differential partonic cross section for channel $\delta$. Throughout this paper we distinguish between the integration variable $\varrho$ and the jet-mass veto $\rho$, with $\varrho \leq \rho$. The integrated distribution, summed over all contributing channels, is then
\begin{equation}
\Sigma(\rho) = \sum_\delta \int \mathrm{d}\mathcal{B}_\delta\,\frac{\mathrm{d}\Sigma_\delta(\rho)}{\mathrm{d}\mathcal{B}_\delta}\,\Xi_\mathcal{B},
\label{eq:Sigma_total}
\end{equation}
where $\Xi_\mathcal{B}$ represents a set of Born-level kinematic cuts. In the region of small jet mass, $\rho \ll 1$, the cross section is dominated by logarithmically enhanced contributions and may be written as
\begin{equation}
\frac{\mathrm{d}\Sigma_\delta(\rho)}{\mathrm{d}\mathcal{B}_\delta} = \frac{\mathrm{d}\sigma_{0,\delta}}{\mathrm{d}\mathcal{B}_\delta}\,f_{\mathcal{B},\delta}(\rho)\,\bigl[1 + \mathcal{O}(\alpha_s)\bigr],
\label{eq:Sigma_fB}
\end{equation}
where $\mathrm{d}\sigma_{0,\delta}/\mathrm{d}\mathcal{B}_\delta$ is the Born-level partonic differential cross section and the function $f_{\mathcal{B},\delta}(\rho)$ resums the hierarchy of large logarithms.

For the C/A algorithm, this resummed factor decomposes as~\cite{Banfi:2004yd, Ziani:2021dxr,Khelifa-Kerfa:2024udm,Khelifa-Kerfa:2025jev,Khelifa-Kerfa:2025jzl}
\begin{equation}
f_{\mathcal{B},\delta}(\rho) = f_{\mathcal{B},\delta}^{\mathrm{global}}(\rho)\;\mathcal{S}_\delta(\rho)\;\mathcal{C}_\delta(\rho),
\label{eq:fB_decomp}
\end{equation}
where $f_{\mathcal{B},\delta}^{\mathrm{global}}(\rho)$ is the algorithm-independent Sudakov form factor arising from the exponentiation of a single soft-gluon emission --- accounting for soft-collinear, hard-collinear, and soft-wide-angle primary radiations from all incoming and outgoing hard partons. This global factor has been computed at NLL accuracy for the same class of observables in~\cite{Dasgupta:2012hg,Ziani:2021dxr} and will not be re-derived here; we refer the reader to those works for explicit expressions.

The two remaining factors in~\eqref{eq:fB_decomp} are algorithm-dependent and constitute the main objects of study in this paper:
\begin{itemize}
\item $\mathcal{S}_\delta(\rho)$ encodes the resummation of leading non-global logarithms arising from correlated secondary gluon emissions, modulated by the C/A clustering constraints on the available phase space.
\item $\mathcal{C}_\delta(\rho)$ captures the resummation of clustering logarithms generated by Abelian, independent primary emissions that are reshuffled into or out of the jet by the C/A recombination procedure.
\end{itemize}
No closed-form analytical all-orders expressions are available for either $\mathcal{S}_\delta$ or $\mathcal{C}_\delta$ with C/A clustering; the present paper provides the first systematic fixed-order expansion of both quantities through four-loop order in the hadronic environment. We expand
\begin{equation}
f_{\mathcal{B},\delta}(\rho) = 1 + f_{\mathcal{B},\delta}^{(1)}(\rho) + f_{\mathcal{B},\delta}^{(2)}(\rho) + f_{\mathcal{B},\delta}^{(3)}(\rho) + f_{\mathcal{B},\delta}^{(4)}(\rho) + \cdots,
\label{eq:fB_expand}
\end{equation}
where $f_{\mathcal{B},\delta}^{(n)}(\rho)$ denotes the $n$-loop contribution of order $\alpha_s^n$ and receives separate CL and NGL pieces,
\begin{equation}
f_{\mathcal{B},\delta}^{(n)}(\rho) \supset \mathcal{C}_{n,\delta}(\rho) + \mathcal{S}_{n,\delta}(\rho), \qquad n \geq 2.
\label{eq:fn_CLs_NGLs}
\end{equation}
The $m$-loop contribution to the jet-mass fraction is expressed via the measurement operator $\hat{\mathcal{U}}_m$, first introduced in~\cite{Schwartz:2014wha} and developed further in~\cite{Khelifa-Kerfa:2015mma,Khelifa-Kerfa:2024roc}, as
\begin{align}
f_{\mathcal{B},\delta}^{(m)}(\rho) &= \sum_X \int_{\xi_1 > \xi_2 > \cdots > \xi_m} \!\!\left(\prod_{i=1}^m \mathrm{d}\Phi_i\right) \hat{\mathcal{U}}_m\,\mathcal{W}_{1\cdots m,\delta}^X \notag \\
&\quad\times \Xi_m^{(\mathrm{C/A})}(k_1,\ldots,k_m),
\label{eq:fm_master}
\end{align}
where $\mathcal{W}_{1\cdots m,\delta}^X$ is the eikonal amplitude squared for $m$ strongly energy-ordered soft gluons in configuration $X$ for channel $\delta$~\cite{Khelifa-Kerfa:2020nlc,Delenda:2015tbo}, the phase-space element for gluon $i$ is
\begin{equation}
\mathrm{d}\Phi_i = \bar{\alpha}_s\frac{\mathrm{d}\xi_i}{\xi_i}\,\mathrm{d}\eta_i\frac{\mathrm{d}\phi_i}{2\pi} = \bar{\alpha}_s\frac{\mathrm{d}\xi_i}{\xi_i}\,R^2 r_i\,\mathrm{d}r_i\,\frac{\mathrm{d}\theta_i}{2\pi},
\label{eq:phasespace}
\end{equation}
with $\bar{\alpha}_s \equiv \alpha_s/\pi$, and $\Xi_m^{(\mathrm{C/A})}(k_1,\ldots,k_m)$ is the C/A clustering function encoding the phase-space constraints imposed by the algorithm on the $m$-gluon configuration. The measurement operator factorizes at each loop order as $\hat{\mathcal{U}}_m = \prod_{i=1}^m \hat{u}_i$, with
\begin{equation}
\hat{u}_i = 1 - \Theta_i^{\mathrm{R}}\,\Theta_i^\rho\,\Theta_i^{\mathrm{in}},
\label{eq:ui}
\end{equation}
where $\Theta_i^{\mathrm{R}} = 1$ for a real emission and zero otherwise, $\Theta_i^{\mathrm{in}} = \Theta(1 - r_i^2)$ enforces that gluon $i$ lies inside the jet after clustering, and $\Theta_i^\rho = \Theta(\varrho_i - \rho)$ restricts to configurations in which gluon $i$'s contribution to the jet mass exceeds the veto $\rho$.

Strong ordering of the gluon transverse momenta, $\xi_1 \gg \xi_2 \gg \cdots \gg \xi_m$, greatly simplifies the clustering sequence: the softer of any two particles always satisfies the C/A inter-particle distance condition before the beam distance, so the C/A clustering function $\Xi_m^{(\mathrm{C/A})}$ reduces to a product of step functions in the polar variables $(r_i, \theta_i)$. The explicit forms of these functions at three- and four-loop order, together with their interplay with the eikonal amplitudes squared, are derived in the following section.

Explicit details of the Born cross sections, parton distribution functions, and the global Sudakov factor are presented in~\cite{Dasgupta:2012hg,Ziani:2021dxr} and will not be reproduced here. In the following, we concentrate entirely on the perturbative computation of $\mathcal{C}_\delta(\rho)$ and $\mathcal{S}_\delta(\rho)$ up to four-loop order.
%%%%%%%%%%%%%%%%%%%%%%%%%%%%%%%%%%%%%%%%%%%%%%%%%%%%%%%%%%
\section{Fixed-order calculations}
\label{sec:FO}

It is straightforward to verify from the definition of the sequential recombination algorithms --- as presented, for instance, in Ref.~\cite{Cacciari:2011ma} --- that all jet algorithms perform identically at one-loop order. Eq.~\eqref{eq:fm_master} gives, up to NLL accuracy~\cite{Ziani:2021dxr,Khelifa-Kerfa:2024udm,Khelifa-Kerfa:2025jev},
\begin{align}
f_{\cB,\delta}^{(1)}(\rho) &= -\asb \Bigg[ \left(\cC_{aj} + \cC_{bj}\right) \frac{L^2}{4} \notag \\
&\quad + \left( \cC_{ab} \frac{R^2}{2} + \left(\cC_{aj} + \cC_{bj}\right) h(R)\right) L \Bigg],
\label{eq:f1}
\end{align}
where $L \equiv \ln(R^2/\rho)$, $h(R) = R^2/8 + R^4/576 + \cO(R^6)$, and the color Casimir scalars are defined in Eq.~\eqref{eq:dipolecolor}. The exponentiation of this one-loop result, after accounting for the running coupling, produces the resummed Sudakov form factor $f_{\cB,\delta}^{\mathrm{global}}(\rho)$ discussed in the paragraph following Eq.~\eqref{eq:fB_decomp}; explicit expressions can be found in Refs.~\cite{Dasgupta:2012hg,Ziani:2021dxr}.

At two-loop order, the $k_t$ and C/A algorithms perform identically, but differently from the anti-$k_t$ algorithm. For completeness, we recall here the two-loop results for the $k_t$ case from our previous work~\cite{Khelifa-Kerfa:2025jev}. The invariant jet-mass distribution at this order takes the form
\begin{align}
f_{\cB,\delta}^{(2)}(\rho) &= \frac{1}{2!}\left[f_{\cB,\delta}^{(1)}(\rho)\right]^2 + \cC_{2,\delta}(\rho) + \cS_{2,\delta}(\rho),
\label{eq:f2}
\end{align}
where the last two terms are the two-loop CL and NGL contributions, respectively. They take the form
\begin{align}
\cC_{2,\delta}(\rho) &= \phantom{-}\frac{1}{2!}\,\asb^2\,L^2\,\cF_{2,\delta}(R), \notag \\
\cS_{2,\delta}(\rho) &= -\frac{1}{2!}\,\asb^2\,L^2\,\cG_{2,\delta}(R),
\label{eq:CL_NGL_2loop}
\end{align}
where the $R$-dependent two-loop coefficients $\cF_{2,\delta}(R)$ and $\cG_{2,\delta}(R)$ are given as power series in $R$ in Eqs.~(3.23) and (3.28) of~\cite{Khelifa-Kerfa:2025jev} for all three partonic channels. The first distinction between the $k_t$ and C/A algorithms arises at three-loop order, which is the subject of the next section.

%%%%%%%%%%%%%%%%%%%%%%%%%%%%%
\subsection{Three-loop order}
\label{sec:3loop}

The three-loop eikonal amplitude, summed over all possible gluon configurations $X$ for a specific partonic channel $\delta$, can be deduced from that of $e^+e^-$ annihilation processes, presented in detail in our previous paper~\cite{Khelifa-Kerfa:2025jzl}, following the procedure outlined in Ref.~\cite{Khelifa-Kerfa:2020nlc}. It reads:
\begin{align}
\sum_{\X} \Uh_3\,\cW_{123,\delta}^{\X,\caa} &= \sum_{\X} \Uh_3\,\cW_{123,\delta}^{\X,\ktt} \notag \\
&\quad - \left(\prod_{i=1}^3 \Theta_i^\rho\right) \Theta_1^\out \Theta_3^\inn\,\O_{12}\,\O_{23}\,\Ob_{13}\,\Delta_{123} \notag \\
&\qquad \times \left[\cW_{123,\delta}^{\V\V\R} + \cW_{123,\delta}^{\R\V\R} + \cW_{123,\delta}^{\R\R\R}\right],
\label{eq:3loop:uWX}
\end{align}
where the first term on the r.h.s.\ is the result for the $k_t$ jet algorithm, reported in Eq.~(3.29) of~\cite{Khelifa-Kerfa:2025jev}, $\O_{ij} = \Theta(d_{jB} - d_{ij})$, and $\Delta_{ijk} = \Theta(d_{jk} - d_{ij})$ with $i < j$. When the constraint $\O_{ij} = 1$, the pair of gluons $(i,j)$ is clustered first into a pseudo-jet with a four-momentum that lies along that of the harder of the two particles. If, instead, $\O_{ij} = 0$ --- and hence $\Ob_{ij} \equiv 1 - \O_{ij} = 1$ --- the pair of gluons $(i,j)$ is not clustered together and each remains at its initial location prior to the application of the C/A algorithm. In both cases, the phase space is nonetheless modified by the algorithm relative to its unclustered form.

When $\Delta_{ijk} = 1$, the pair of particles $(i,j)$ is clustered first, having the smallest distance. The opposite case, $\Delta_{ijk} = 0$, corresponds to the pair $(j,k)$ being clustered first. If, for example, both $\Delta_{123}$ and $\Delta_{231}$ are equal to unity, there are two competing small distances, $d_{12}$ and $d_{23}$, that must be manually ordered for the analytical calculation to cover the full available phase space: (a)~$d_{12} < d_{23}$: gluons $k_1$ and $k_2$ are clustered first, and by strong energy ordering the resulting pseudo-jet lies along the harder gluon $k_1$; gluon $k_3$ is then left unclustered; (b)~$d_{23} < d_{12}$: gluons $k_2$ and $k_3$ are clustered first into a pseudo-jet along the harder gluon $k_2$, which is in turn clustered with $k_1$; gluon $k_1$ is then left unclustered. This simple example illustrates how the ordering of the distance measures can alter the final gluon configuration, and hence each gluon's contribution to the observable distribution. Such an ordering requirement is absent for both the anti-$k_t$ and $k_t$ algorithms, since their distance measures are automatically ordered by the transverse momenta that already appear in their definitions.

It is worth noting that, in Eq.~\eqref{eq:3loop:uWX}, the location of gluon $k_2$ is not explicitly shown and therefore does not play a decisive role in the C/A clustering correction. This can be explained straightforwardly as follows: from the product of the constraints $\O_{12}$ and $\O_{23}$ it is evident that gluon $k_2$ will always be clustered either with the hardest gluon $k_1$ or the softest gluon $k_3$. In the first case, $d_{12} < d_{23}$, $k_2$ is dragged by $k_1$ regardless of its initial position (in or out of the jet), and the resulting pseudo-jet is essentially $k_1$ by strong energy ordering; one is then left with only gluons $k_1$ and $k_3$ to determine whether the configuration contributes to the jet mass. In the second case, $d_{12} > d_{23}$, $k_2$ drags $k_3$ into a pseudo-jet that lies essentially along $k_2$, which is itself then dragged by the harder gluon $k_1$, yielding a final pseudo-jet along $k_1$. In both cases, $k_2$ plays the role of transporting $k_3$ toward or away from $k_1$ depending on their relative angular separation --- a condition that is independent of whether $k_2$ lies inside or outside the jet.

In our recent paper on C/A clustering in $e^+e^-$~\cite{Khelifa-Kerfa:2025jzl}, we discussed in full detail how the C/A clustering proceeds at three-loop order. We listed all possible scenarios and orderings that must be taken into account and combined them to arrive at the analog of Eq.~\eqref{eq:3loop:uWX} for the $e^+e^-$ case. The calculations for $V/H$+jet follow exactly the same procedure for all three channels, the only difference being in the corresponding Born color factors.

Substituting the explicit formulas for the eikonal amplitudes squared from Ref.~\cite{Khelifa-Kerfa:2020nlc}, one obtains, for the sum in the square bracket in Eq.~\eqref{eq:3loop:uWX},
\begin{widetext}
\begin{align}
&\prod_{\sigma=1}^3 \left(\sum_{(i_\sigma j_\sigma)\in\Delta_\delta} \cC_{i_\sigma j_\sigma} w_{i_\sigma j_\sigma}^\sigma\right) + \sum_{\substack{k,\ell,m=1\\k\neq\ell\neq m}}^3 \left(\CA \sum_{(ij)\in\Delta_\delta} \cC_{ij}\,\cA_{ij}^{k\ell}\right)\!\left(\sum_{(i'j')\in\Delta_\delta} \cC_{i'j'} w_{i'j'}^m\right) \notag \\
&\quad + \CAsq \sum_{(ij)\in\Delta_\delta} \cC_{ij}\!\left(\cA_{ij}^{12}\,\cAb_{ij}^{13} + \cB_{ij}^{123}\right)
+ \cQ_\delta \sum_{(ijk)\in\pi_\delta}\!\left(\cG_{ij}^{k1}(2,3) + 2\leftrightarrow 3\right),
\label{eq:3loop:uWX-b}
\end{align}
\end{widetext}
where the one-loop antenna function is given in Eq.~\eqref{eq:antenna}, the set of all dipoles in channel $\delta$ is $\Delta_\delta = \{(ab),(aj),(bj)\}$, the set of triplets in channel $\delta$ is $\pi_\delta = \{(abj),(ajb),(bja)\}$, the color Casimir scalars are defined in Eq.~\eqref{eq:dipolecolor}, and the quadrupole color factor, $\cQ_{\delta}$, takes the channel-dependent values~\cite{Khelifa-Kerfa:2020nlc}
\begin{equation}
\cQ_{\delta_1} = \cQ_{\delta_2} = \CAsq\!\left(\CA - 2\CF\right), \qquad \cQ_{\delta_3} = 6\,\CA.
\label{eq:QuadColor}
\end{equation}
The higher-order antenna functions are defined as follows: \cite{Delenda:2015tbo, Khelifa-Kerfa:2020nlc}
\begin{subequations}
\label{eq:3loop:DipoleFunctions}
\begin{align}
\cA_{\alpha\beta}^{ij} &= w_{\alpha\beta}^i\!\left(w_{\alpha i}^j + w_{i\beta}^j - w_{\alpha\beta}^j\right), \\
\cB_{\alpha\beta}^{ijk} &= w_{\alpha\beta}^i\!\left(\cA_{\alpha i}^{jk} + \cA_{i\beta}^{jk} - \cA_{\alpha\beta}^{jk}\right), \\
\cG_{\alpha\beta}^{ij}(m,n) &= w_{\alpha\beta}^j\,T_{\alpha\beta}^{ij}(n)\,U_{\alpha\beta}^{ij}(m),
\end{align}
with the cross-channel functions
\begin{align}
T_{\alpha\beta}^{ij}(n) &= w_{\alpha\beta}^n + w_{ij}^n - w_{\alpha i}^n - w_{\beta j}^n, \\
U_{\alpha\beta}^{ij}(n) &= w_{\alpha\beta}^n + w_{ij}^n - w_{\alpha j}^n - w_{\beta i}^n.
\end{align}
\end{subequations}
The first term in Eq.~\eqref{eq:3loop:uWX-b} is the source of CLs, since it arises from independent primary emissions whose real-virtual cancellation is disturbed by the clustering, whereas the remaining three terms give rise to NGLs. Upon integration, one can show that the perturbative distribution at this loop order takes the form---analogous to the $k_t$ case~\cite{Khelifa-Kerfa:2025jev}---
\begin{align}
f_{\cB,\delta}^{(3),\caa}(\rho) &= \frac{1}{3!}\!\left[f_{\cB,\delta}^{(1)}\right]^3 + f_{\cB,\delta}^{(1)}\times \left[\cC_{2,\delta} + \cS_{2,\delta}\right] \notag \\
&\quad + \cC_{3,\delta}^{\caa} + \cS_{3,\delta}^{\caa}.
\label{eq:3loop:fb-FinalForm}
\end{align}
The first two terms on the r.h.s.\ of Eq.~\eqref{eq:3loop:fb-FinalForm} have already been computed and are identical to the $k_t$ case. The last two terms, $\cC_{3,\delta}^{\caa}$ and $\cS_{3,\delta}^{\caa}$, are the new contributions arising from the application of C/A clustering to the invariant jet-mass distribution at three-loop order. We discuss them in detail below.

%%%%%%%%%%%%%%%%%%%%%%%%%%
\subsubsection{CLs}

\begin{figure}[t]
\centering
\includegraphics[scale=0.55]{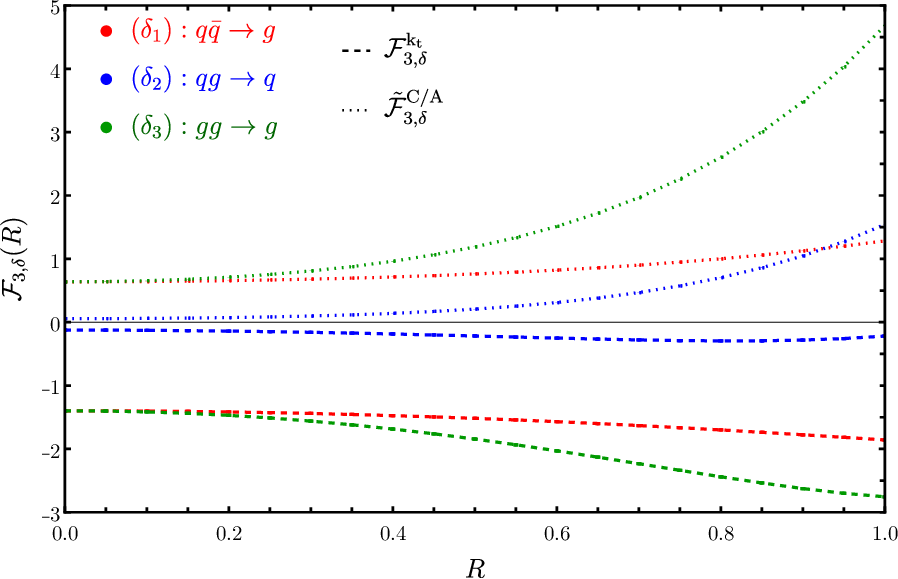}
\caption{The pure C/A corrections to the CL coefficients at three-loop order as a function of the jet radius $R$ (dotted curves) for all three Born channels. The dashed curves correspond to the CL coefficients in the $k_t$ clustering case.}
\label{fig:F3a}
\end{figure}

The primary-emission piece in Eq.~\eqref{eq:3loop:uWX-b} gives rise to clustering logarithms and may be cast in the usual factorized form~\cite{Dasgupta:2012hg,Kerfa:2012yae,Khelifa-Kerfa:2011quw,Khelifa-Kerfa:2015mma,Ziani:2021dxr}:
\begin{subequations}
\begin{align}
\mathcal{C}_{3,\delta}^{\mathrm{C/A}}(\rho) &= -\frac{1}{3!}\,\bar{\alpha}_s^3\,L^3\,\mathcal{F}_{3,\delta}^{\mathrm{C/A}}(R),
\label{eq:3loop:CLs}
\end{align}
where the full C/A coefficient for CLs at three-loop order decomposes as
\begin{align}
\mathcal{F}_{3,\delta}^{\mathrm{C/A}} = \mathcal{F}_{3,\delta}^{k_t} + \widetilde{\mathcal{F}}_{3,\delta}^{\,\mathrm{C/A}},
\label{eq:3loop:F3-decomp}
\end{align}
where $\mathcal{F}_{3,\delta}^{k_t}$ is the contribution already present in $k_t$ clustering. Recall that the C/A algorithm strictly encompasses $k_t$: every configuration accounted for by $k_t$ clustering is automatically included in C/A clustering, with C/A generating additional contributions not captured by $k_t$. This has been discussed extensively in Ref.~\cite{Khelifa-Kerfa:2025jzl}. The pure C/A correction at this order, $\widetilde{\mathcal{F}}_{3,\delta}^{\,\mathrm{C/A}}$, reads:
\begin{align}
\widetilde{\mathcal{F}}_{3,\delta}^{\,\mathrm{C/A}}(R) &= R^6 \int_{1_{\mathrm{out}}} \left[\int_{2_{\mathrm{in}}} + \int_{2_{\mathrm{out}}}\right] \int_{3_{\mathrm{in}}} \Omega_{12}\,\Omega_{23}\,\bar{\Omega}_{13}\,\Delta_{123} \notag \\
&\phantom{{}= R^6}\times \left[\prod_{\sigma=1}^{3} \sum_{(i_\sigma j_\sigma)\in\Delta_\delta} \mathcal{C}_{i_\sigma j_\sigma}\,w_{i_\sigma j_\sigma}^{\sigma}\right],
\label{eq:3loop:F3-A}
\end{align}
\end{subequations}
where the following integral shorthand notation is adopted~\cite{Khelifa-Kerfa:2025jev}:
\begin{align}
\int_{i_{\mathrm{in}}} &\equiv \int_0^1 r_i\,\mathrm{d}r_i \int_0^{2\pi}\frac{\mathrm{d}\theta_i}{2\pi}, \notag \\
\int_{i_{\mathrm{out}}} &\equiv \int_1^{\pi/|R\sin\theta_i|} r_i\,\mathrm{d}r_i \int_0^{2\pi}\frac{\mathrm{d}\theta_i}{2\pi},
\end{align}
where the upper limit on the second integral has been discussed in our previous works (see, for instance, the paragraph following Eq.~(3.26) of Ref.~\cite{Khelifa-Kerfa:2025jev}). To account for integration over all positions of gluon $k_2$, we exploit the completeness relation $\int_{2_{\mathrm{in}}}+\int_{2_{\mathrm{out}}}=1$. Both integrals are evaluated numerically using the multidimensional \texttt{Cuba} library~\cite{Hahn:2004fe} and summed to yield the results displayed in Fig.~\ref{fig:F3a}. The numerical values of  $\mathcal{F}_{3,\delta}^{\ktt}$ and  $\mathcal{F}_{3,\delta}^{\caa}$ at some representative values of $R$ are shown in Table \ref{tab:F3G3-Values}.

Figure~\ref{fig:F3a} compares $\widetilde{\mathcal{F}}_{3,\delta}^{\,\mathrm{C/A}}$ with $\mathcal{F}_{3,\delta}^{k_t}$ for all three Born channels as a function of the jet radius $R$. The pure C/A corrections are generally of opposite sign to the $k_t$ coefficients and smaller in magnitude; they thus serve to reduce the impact of the $k_t$ contributions.

\begin{figure}[t]
\centering
\includegraphics[scale=0.55]{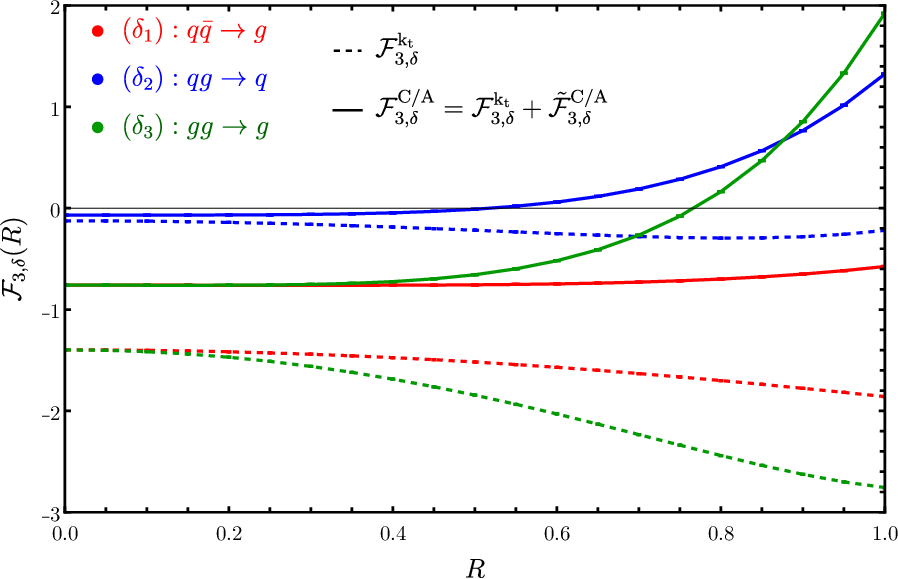}
\caption{Comparison of the full C/A (solid) and $k_t$ (dashed) CL coefficients at three-loop order as a function of $R$ for all three Born channels.}
\label{fig:F3b}
\end{figure}

In Fig.~\ref{fig:F3b} we plot the full C/A CL coefficient alongside the $k_t$ one. As expected, combining the $k_t$ contribution with the pure C/A correction $\widetilde{\mathcal{F}}_{3,\delta}^{\,\mathrm{C/A}}$ yields a full C/A CL coefficient that is smaller in magnitude than its $k_t$ counterpart. This is an advantageous feature of the C/A algorithm, since it is phenomenologically desirable to minimize or entirely eliminate the contribution of CLs to the observable distribution. Several important observations follow from Fig.~\ref{fig:F3b}:
\begin{itemize}
\item Unlike the CL coefficients for $k_t$ clustering, which are negative for all values of $R$, those of C/A change sign across the range of $R$. This is especially true for channels $(\delta_2)$ and $(\delta_3)$. It follows from Eq.~\eqref{eq:3loop:CLs} that the CL  contribution to the jet-mass distribution in $k_t$ clustering is positive across the whole range of $R$ and for all Born channels. A similar behavior is seen for channel $(\delta_1)$ in C/A clustering. By contrast, channels $(\delta_2)$ and $(\delta_3)$ do not maintain a constant sign over all values of $R$. Specifically, channel $(\delta_2)$ has a vanishing CL contribution at $R\sim 0.5$, while channel $(\delta_3)$ has a vanishing CL contribution at $R\sim 0.78$. These particular values of $R$ could in principle be exploited to eliminate the CL contribution entirely. However, one must also verify the magnitude of the NGLs at such values, since the ideal scenario would be for both CLs and NGLs to be simultaneously suppressed or to mutually cancel.

\item For small values of $R$, the CL coefficients for channels $(\delta_1)$ and $(\delta_3)$ are identical in size, owing to the equality of their color factors --- both channels possess an outgoing hard gluon in the final state. They do, however, become distinct at larger values of $R$, with the divergence beginning around $R\sim 0.35$ for C/A and $R\sim 0.15$ for $k_t$.

\item The \textit{edge (boundary)} effect is evident for both clustering algorithms. Specifically, as the jet radius shrinks to zero, the CL coefficients do not tend to zero but instead approach a finite constant. This behavior has been explained in detail in our previous papers (see, for instance, Refs.~\cite{Kerfa:2012yae,Khelifa-Kerfa:2011quw,Khelifa-Kerfa:2024udm}).

\item In the limit $R\to 0$, one recovers the $e^+e^-$ findings of Ref.~\cite{Khelifa-Kerfa:2025jzl}:
\begin{align}
&\lim_{R\to 0}\mathcal{F}_{3,\delta_1}^{\mathrm{C/A}}(R) = \lim_{R\to 0}\mathcal{F}_{3,\delta_3}^{\mathrm{C/A}}(R) = -0.76 \notag\\
&  \hfill \hspace{4.7cm} \simeq -0.028\,\CAcub, \notag\\
&\lim_{R\to 0}\mathcal{F}_{3,\delta_2}^{\mathrm{C/A}}(R) = -0.068 \simeq -0.028\, \CFcub.
\label{eq:3loop:F3-R=0}
\end{align}

\item The $R$-dependence of the CL coefficients in C/A clustering differs among the three channels, with channel $(\delta_3)$ exhibiting the strongest $R$-dependence and channel $(\delta_1)$ the weakest. Moreover, the $R$-dependence of the CL coefficients is more pronounced in C/A clustering than in $k_t$ clustering. This may be attributed to the fact that the C/A algorithm admits more gluon configurations that contribute to the jet-mass observable than $k_t$; furthermore, such configurations depend on $R$ through the angular distance measure.

\item Quantitatively, the CL coefficients in C/A clustering are reduced by approximately 45\% relative to the $k_t$ case for all three channels at $R=0$. For channel $(\delta_1)$, this reduction grows steadily, reaching approximately 70\% at $R=1$. For channels $(\delta_2)$ and $(\delta_3)$, the reduction reaches 100\% at the values of $R$ where the C/A CL coefficients vanish. For channel $(\delta_2)$, the C/A CL coefficient subsequently changes sign and grows to exceed the magnitude of its $k_t$ counterpart for $R\gtrsim 0.8$. For channel $(\delta_3)$, the reduction diminishes to approximately 30\% at $R\sim 1$, with the C/A coefficient carrying the opposite sign to that of $k_t$.
\end{itemize}

\begin{table}[ht]
\centering
\caption{Numerical values of the three-loop CL and NGL coefficients, $\cF_{3, \delta}$ and $\cG_{3, \delta}$, for the $k_t$ and C/A   algorithms at representative jet radii $R$ for all three Born channels. }
\label{tab:F3G3-Values}
\setlength{\tabcolsep}{7pt}
\renewcommand{\arraystretch}{1.1}
\begin{tabular}{cccccc}
\toprule
\textbf{$R$} & \textbf{Channels} & $\bm{\cF_{3, \delta}^{\ktt}}$ & $\bm{\cF_{3, \delta}^{\caa}}$ & $\bm{\cG_{3, \delta}^{\ktt}}$ & $\bm{\cG_{3, \delta}^{\caa}}$ \\
%\midrule
%\multirow{3}{*}{$0$} & $(\delta_1)$ & $-1.40$ & $-0.76$ & $24.08$ & $5.95$ \\
%                   & $(\delta_2)$ & $-0.12$ & $-0.07$ & $7.57$ & $1.54$ \\
%                   & $(\delta_3)$ & $-1.40$ & $-0.76$ & $24.08$ & $5.95$ \\
\midrule
\multirow{3}{*}{$0.4$} & $(\delta_1)$ & $-1.47$ & $-0.76$ & $23.81$ & $5.06$ \\
                   & $(\delta_2)$ & $-0.19$ & $-0.05$ & $13.16$ & $4.94$ \\
                   & $(\delta_3)$ & $-1.69$ & $-0.72$ & $28.86$ & $7.04$ \\
\midrule
\multirow{3}{*}{$0.6$} & $(\delta_1)$ & $-1.57$ & $-0.75$ & $24.66$ & $5.02$ \\
                   & $(\delta_2)$ & $-0.25$ & $-0.06$ & $8.08$ & $-3.05$ \\
                   & $(\delta_3)$ & $-2.03$ & $-0.52$ & $23.63$ & $-2.97$ \\
\midrule
\multirow{3}{*}{$0.7$} & $(\delta_1)$ & $-1.63$ & $-0.73$ & $25.32$ & $5.09$ \\
                   & $(\delta_2)$ & $-0.28$ & $-0.19$ & $3.40$ & $-9.69$ \\
                   & $(\delta_3)$ & $-2.23$ & $-0.27$ & $19.25$ & $-10.67$ \\
\midrule
\multirow{3}{*}{$0.8$} & $(\delta_1)$ & $-1.70$ & $-0.70$ & $26.10$ & $5.20$ \\
                   & $(\delta_2)$ & $-0.28$ & $-0.29$ & $0.41$ & $-17.82$ \\
                   & $(\delta_3)$ & $-2.44$ & $0.16$ & $14.23$ & $-19.61$ \\
\midrule
\multirow{3}{*}{$1.0$} & $(\delta_1)$ & $-1.85$ & $-0.57$ & $27.73$ & $5.13$ \\
                   & $(\delta_2)$ & $-0.22$ & $1.32$ & $-15.49$ & $-36.99$ \\
                   & $(\delta_3)$ & $-2.76$ & $1.92$ & $4.95$ & $-38.96$ \\
\bottomrule
\end{tabular}
\end{table}

In the following section, we discuss the NGL contribution to the jet-mass distribution at three-loop order.

%%%%%%%%%%%%%%%%%%%%%%
\subsubsection{NGLs}

\begin{figure}[t]
\centering
\includegraphics[scale=0.55]{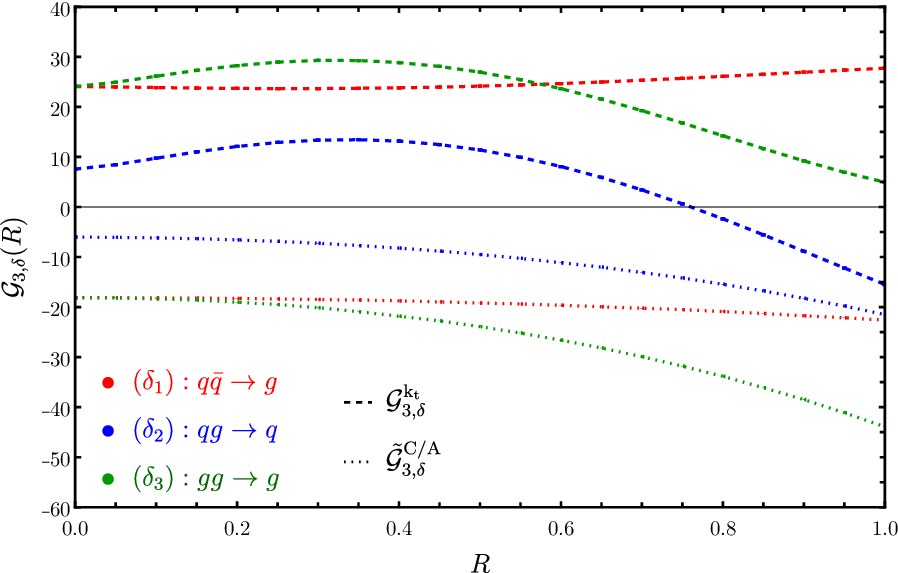}
\caption{The pure C/A corrections to the NGL coefficients at three-loop order as a function of the jet radius $R$ (dotted curves) for all three Born channels. The dashed curves correspond to the NGL coefficients in the $k_t$ clustering case.}
\label{fig:G3a}
\end{figure}

The computation of NGL coefficients follows an analogous procedure to that of CLs. In Eq.~\eqref{eq:3loop:uWX-b} there are three terms contributing to NGLs at this loop order. The first term, proportional to $\mathcal{A}_{ij}^{k\ell}\times w_{i'j'}^m$, arises from interference between the one- and two-loop perturbative distributions. The second, proportional to $\mathcal{A}_{ij}^{12}\bar{\mathcal{A}}_{ij}^{13}$ and $\mathcal{B}_{ij}^{123}$, corresponds to two-loop dipole-dipole interference and a genuine three-loop dipole contribution. The last term, proportional to $\mathcal{G}_{ij}^{k\ell}$, corresponds to the quadrupole (ghost) contribution (see Ref.~\cite{Khelifa-Kerfa:2020nlc} for full details). The total NGL contribution from all three contributions can be shown to take the usual form:
\begin{align}
\mathcal{S}_{3,\delta}^{\mathrm{C/A}}(\rho) = +\frac{1}{3!}\,\bar{\alpha}_s^3\,L^3\,\mathcal{G}_{3,\delta}^{\mathrm{C/A}}(R),
\end{align}
where the NGL coefficient is given by
\begin{align}
\mathcal{G}_{3,\delta}^{\mathrm{C/A}}(R) = \mathcal{G}_{3,\delta}^{k_t}(R) + \widetilde{\mathcal{G}}_{3,\delta}^{\,\mathrm{C/A}}(R),
\end{align}
with the first term corresponding to the $k_t$ contribution, already computed in Ref.~\cite{Khelifa-Kerfa:2025jev}, and the second term being the pure C/A correction at this loop order. Recall that each of these two terms contains all three aforementioned contributions, namely interference, dipole, and quadrupole. These integrals may be evaluated numerically using, for instance, the \texttt{Cuba} library. The results are plotted in Fig.~\ref{fig:G3a}. Analogously to the CL case, the pure C/A corrections to the NGL distribution are of opposite sign to the $k_t$ ones and of comparably smaller magnitude; they thus serve to reduce the NGL contribution relative to both $k_t$ and anti-$k_t$. The latter usually produces the largest NGL contribution (see, for instance, Refs.~\cite{Khelifa-Kerfa:2024udm,Khelifa-Kerfa:2025jev}). Note that, unlike the $k_t$ case, the C/A corrections maintain a consistent sign across the whole range of $R$ for all three channels.

In Fig.~\ref{fig:G3b} we plot the full C/A NGL coefficients for all three Born channels as a function of the jet radius alongside their $k_t$ counterparts. Several observations follow from Fig.~\ref{fig:G3b}:
\begin{figure}[t]
\centering
\includegraphics[scale=0.55]{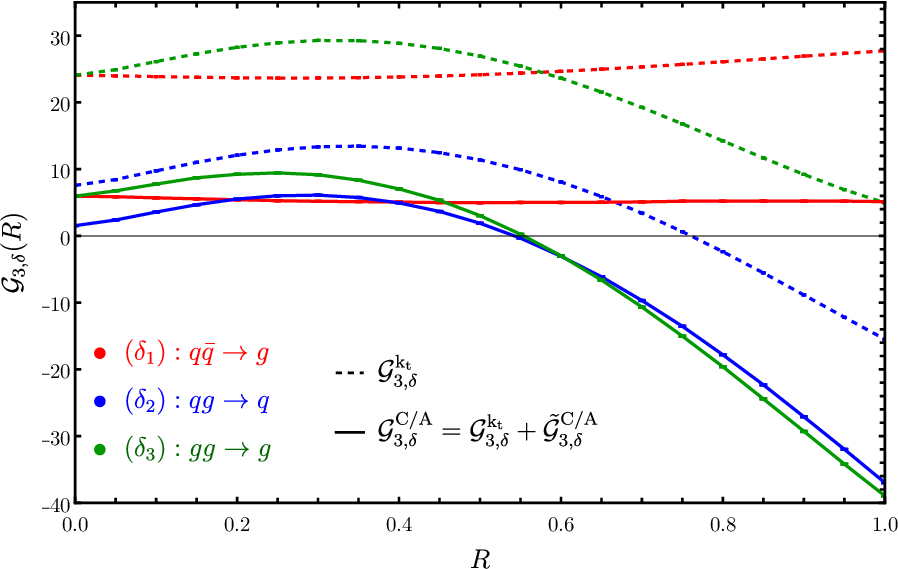}
\caption{Comparison of the full C/A (solid) and $k_t$ (dashed) NGL coefficients at three-loop order as a function of $R$ for all three Born channels.}
\label{fig:G3b}
\end{figure}
\begin{itemize}
\item As anticipated, the NGL coefficients for C/A clustering are generally smaller in magnitude than those for $k_t$. However, this pattern is not consistent over the whole range of $R$. There are intervals of $R$ in which the C/A coefficients exceed those of $k_t$; this occurs after the coefficients change sign.

\item The boundary effect persists across all channels and all jet algorithms.

\item The NGL coefficients in C/A clustering for channels $(\delta_2)$ and $(\delta_3)$ vanish at $R\sim 0.55$. This is to be compared with only one channel for $k_t$ clustering, namely channel $(\delta_2)$, which vanishes at $R\sim 0.78$.

\item The reduction in the magnitude of the NGL coefficients in C/A clustering relative to $k_t$ is approximately 75--79\% for all three Born channels at $R=0$. For channel $(\delta_1)$, this reduction continues steadily, reaching approximately 81\% at $R=1$. Note that for this channel there is no change of sign. For channels $(\delta_2)$ and $(\delta_3)$, the C/A NGL coefficients grow to exceed those of $k_t$ beyond the value of $R$ at which they vanish (i.e., beyond $R\sim 0.55$).

\item In the limit of small $R$, one obtains from Fig.~\ref{fig:G3b}:
\begin{align}
&\lim_{R\to 0}\mathcal{G}_{3,\delta_1}^{\mathrm{C/A}} = \lim_{R\to 0}\mathcal{G}_{3,\delta_3}^{\mathrm{C/A}} = 5.95, \notag\\
&\lim_{R\to 0}\mathcal{G}_{3,\delta_2}^{\mathrm{C/A}} = 1.54.
\end{align}
These values differ considerably from those reported for the $e^+e^-$ dijet-mass observable~\cite{Khelifa-Kerfa:2025jzl}, where $\CFsq \CA (0.43) + \CF \CAsq (0.37)\simeq 6.73$ and $\CAcub (0.43+0.37)\simeq 21.6$. The latter were also found not to match those calculated for the hemisphere-mass observable~\cite{Khelifa-Kerfa:2025cdn}. The small-$R$ limit of the $k_t$ NGL coefficients does coincide with those computed for the $e^+e^-$ dijet-mass observable, as illustrated in Ref.~\cite{Khelifa-Kerfa:2025jev}. They read (see also Fig.~\ref{fig:G3b}):
\begin{align}
& \lim_{R\to 0}\mathcal{G}_{3,\delta_1}^{k_t} = \lim_{R\to 0}\mathcal{G}_{3,\delta_3}^{k_t} = 7.5, \notag\\
& \lim_{R\to 0}\mathcal{G}_{3,\delta_2}^{k_t} = 24.07.
\end{align}
Note that the $e^+e^-$ results in C/A clustering are close to the $k_t$ ones at small $R$ because the pure C/A corrections are quite small in that case. By contrast, the pure C/A corrections in the $V/H$+jet processes studied in this paper are significant, reaching well beyond 70\% for small $R$, as discussed above (see Fig.~\ref{fig:G3a}). This may be attributed to the fact that, according to Eq.~\eqref{eq:3loop:uWX-b}, all eikonal amplitudes squared are summed together as they all share the same phase-space constraint --- unlike the $k_t$ case, in which the eikonal amplitudes squared are multiplied by different phase-space constraints, causing a partial cancellation among them (see Eqs.~(3.48) of Ref.~\cite{Khelifa-Kerfa:2025jev}).
\end{itemize}

\begin{figure}[t]
\centering
\includegraphics[scale=0.55]{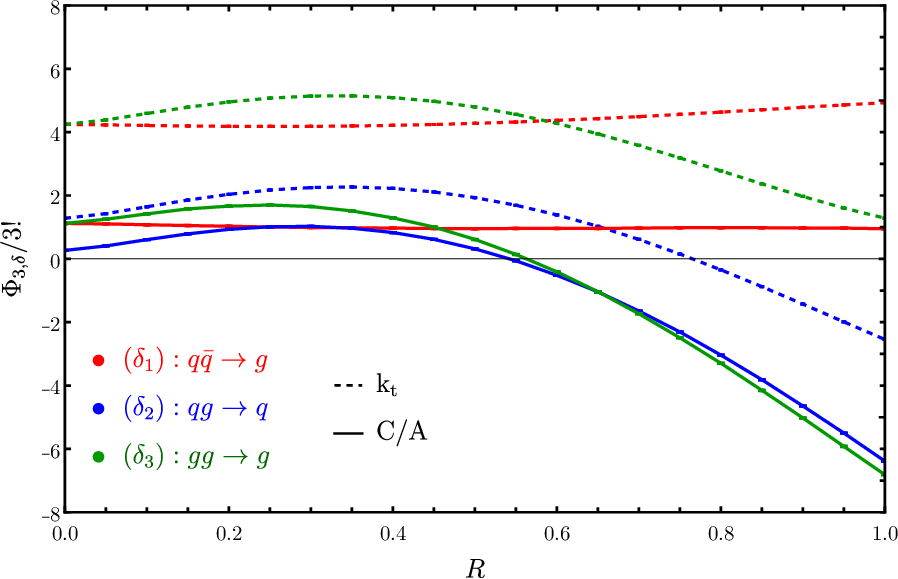}
\caption{Comparison of the sum of CL and NGL coefficients for C/A (solid) and $k_t$ (dashed) clustering at three-loop order as a function of $R$ for all three Born channels.}
\label{fig:GF3}
\end{figure}

To assess the combined impact of both types of non-global logarithms --- CLs and NGLs --- on the distribution of the invariant-mass observable at this loop order, we plot, in Fig.~\ref{fig:GF3},  their sum
\begin{align}
\cC_{3,\delta}(\rho) + \cS_{3,\delta}(\rho) = \frac{1}{3!}\asb^3\, L^3\, \Phi_{3, \delta}(R),
\end{align}
for both $k_t$ and C/A clusterings, where $ \Phi_{3, \delta} \equiv - \cF_{3, \delta} + \cG_{3, \delta}$.
We observe that for $R\lesssim 0.7$ the sum of both coefficients is of order unity or smaller for all three channels in the C/A clustering case. In fact, for channels $(\delta_2)$ and $(\delta_3)$ the combined non-global effect vanishes for $R\sim 0.55$--$0.58$. A similar behavior is seen for $k_t$, but only for channel $(\delta_2)$ in the range $R\in[0,1]$. This confirms the conclusion of Ref.~\cite{Khelifa-Kerfa:2025jzl} for $e^+e^-$ processes that the C/A jet algorithm is preferred to $k_t$ --- and over anti-$k_t$ --- in mitigating the impact of large non-global logarithms over a wide range of $R$, and in eliminating them entirely for specific values of $R$. This conclusion holds, however, only at three-loop order. One must verify it at higher loop orders as well as at all orders. The former task is carried out in the next section, in which we address the impact of the said logarithms at four loops.

%%%%%%%%%%%%%%%%%%%%%%%%%%%%%%%%%%%%%%%%%
\subsection{Four-loop order}
\label{sec:4loop}

The four-loop calculation of the invariant jet-mass distribution for the leading high-$p_t$ jet follows an identical strategy to the three-loop case presented in the previous section. The details are based on the results of Refs.~\cite{Khelifa-Kerfa:2020nlc,Khelifa-Kerfa:2024udm,Khelifa-Kerfa:2025jev,Khelifa-Kerfa:2025jzl}. The full structure of the C/A integrand is identical to that computed for the $e^+e^-$ case in~\cite{Khelifa-Kerfa:2025jzl}, the only difference being in the eikonal amplitudes squared, which are instead taken from~\cite{Khelifa-Kerfa:2020nlc}. The main steps leading to the final results are as follows:
\begin{itemize}
\item One enumerates all initial in-jet and out-of-jet gluon configurations at four loops. These include configurations --- prior to applying the C/A algorithm --- in which only gluon $k_4$ is inside the jet and all others are outside; gluons $k_3$ and $k_4$ are inside and the remaining two are outside; and so on. For each such configuration one considers all possible pairwise clusterings among the four gluons, characterized by the constraints $\Delta_{12j}$, $\Delta_{13j}$, $\Delta_{14j}$, $\Delta_{23j}$, $\Delta_{24j}$, and $\Delta_{34j}$. Each clustering constraint of the form $\Delta_{\alpha\beta j}$, where $j$ denotes the leading jet, takes the value $1$ or $0$. If it equals $1$, gluons $k_\alpha$ and $k_\beta$ are clustered first, i.e., they have the smallest pairwise distance measure. If it equals $0$, the smallest distance is that between the softer gluon $k_\beta$ and the jet, and the two gluons are not clustered together. There is a total of $2^6 = 64$ such configurations to be considered. Full details are given in Ref.~\cite{Khelifa-Kerfa:2025jzl}.

\item For configurations in which more than one $\Delta_{\alpha\beta j}$ equals $1$, all possible orderings of the smallest distances must be considered. For example, if $\Delta_{12j} = \Delta_{13j} = 1$, one must consider both $d_{12} > d_{13}$ and $d_{12} < d_{13}$. There is a total of $1957$ such orderings across the $64$ configurations.

\item For each of the above gluon configurations there are $16$ possible real/virtual assignments, corresponding to each of the four gluons being either real~(R) or virtual~(V). Of these, only $8$ need to be computed explicitly; the remaining $8$ follow by symmetry \cite{Khelifa-Kerfa:2015mma, Khelifa-Kerfa:2024udm}. The $8$ independent configurations are: RRRR, RRVR, RVRR, VRRR, RVVR, VRVR, VVRV, and VVVR. Here, for instance, RRRR denotes all four gluons real, and RVVR denotes gluons $k_1$ and $k_4$ real with $k_2$ and $k_3$ virtual. The explicit formulas for the corresponding eikonal amplitudes squared are given in~\cite{Khelifa-Kerfa:2020nlc}.

\item The C/A algorithm is applied to each configuration to determine the final in-jet or out-of-jet status of each of the four gluons. Only gluons that end up inside the measured jet contribute to its mass. NGLs and CLs arise from configurations in which the real-emission contribution to the jet mass is not fully canceled by the corresponding virtual corrections.

\item Accounting for all of the above configurations is a tedious and complex task that can, nonetheless, be fully automated. To this end, we have developed a \texttt{Python} code to perform the bookkeeping, together with a \texttt{Mathematica} script to organize the output expressions. Since the C/A clustering contains the $k_t$ clustering as a subset, the \texttt{Python} code is written to omit contributions already accounted for in the $k_t$ case, so that the final output consists of the pure new C/A corrections. The $k_t$ contributions have already been computed in our previous work~\cite{Khelifa-Kerfa:2025jev}.

\item The resulting pure C/A expressions constitute the integrand to be evaluated numerically. Of all the phase-space regions enumerated above, only four produce non-vanishing real-virtual mis-cancellations and hence contribute to the jet-mass distribution, namely: $1_\out 2_\out 3_\out 4_\inn$, $1_\out 2_\out 3_\inn 4_\inn$, $1_\out 2_\inn 3_\out 4_\inn$, and $1_\out 2_\inn 3_\inn 4_\inn$.
\end{itemize}
Following the above analysis, one can show that the invariant jet-mass distribution at four-loop order takes an analogous form to those at two- and three-loop order, Eqs.~\eqref{eq:f2} and~\eqref{eq:3loop:fb-FinalForm} (see also Refs.~\cite{Khelifa-Kerfa:2025jev,Khelifa-Kerfa:2025jzl}):
\begin{align}
f_{\cB,\delta}^{(4),\caa}(\rho) &= \frac{1}{4!}\!\left[f_{\cB,\delta}^{(1)}\right]^4 + f_{\cB,\delta}^{(1)}\times \left[\cC_{3,\delta}^{\caa} + \cS_{3,\delta}^{\caa}\right] \notag \\
&\quad + \frac{1}{2!}\!\left[f_{\cB,\delta}^{(1)}\right]^2\times \left[\cC_{2,\delta} + \cS_{2,\delta} \right] + \frac{1}{2!}\!\left[\cC_{2,\delta} \right]^2 \notag \\
&\quad + \frac{1}{2!}\!\left[\cS_{2,\delta} \right]^2 +  \cC_{2, \delta} \times \cS_{2, \delta} + \cC_{4,\delta}^{\caa} + \cS_{4,\delta}^{\caa}.
\label{eq:4loop:fB}
\end{align}
The last two terms are the new four-loop CL and NGL contributions. They admit the usual factorized form
\begin{subequations}
\label{eq:4loop:CLs-NGLs}
\begin{align}
\cC_{4,\delta}^{\caa}(\rho) &= +\frac{1}{4!}\,\asb^4\,L^4\,\cF_{4,\delta}^{\caa}(R), \\
\cS_{4,\delta}^{\caa}(\rho) &= -\frac{1}{4!}\,\asb^4\,L^4\,\cG_{4,\delta}^{\caa}(R).
\end{align}
\end{subequations}
Although Eqs.~\eqref{eq:4loop:CLs-NGLs} define $\cF_{4,\delta}^{\caa}$ and $\cG_{4,\delta}^{\caa}$ as the formal coefficients of the four-loop CL and NGL contributions separately, the complex interplay among color factors, kinematics, and jet-algorithm constraints at this order prevents their independent numerical determination for C/A clustering. Unlike the situation at two- and three-loop order~--- where the primary-emission piece of the integrand (the source of CLs) and the correlated-emission piece (the source of NGLs) can be isolated term by term~--- the C/A clustering constraints at four loops generate phase-space constraints that couple both sources inseparably.  Only their combination
\begin{align}
\cC_{4,\delta}^{\caa}(\rho) + \cS_{4,\delta}^{\caa}(\rho)   = \frac{1}{4!}\,\asb^4 L^4 \underbrace{\bigl(\cF_{4,\delta}^{\caa}(R) - \cG_{4,\delta}^{\caa}(R)\bigr)}_{\displaystyle\equiv\;\Phi_{4,\delta}^{\caa}(R)} \label{eq:4loop:combined}
\end{align}
is numerically accessible, where we have introduced the shorthand $\Phi_{4,\delta}^{A}(R) \equiv \cF_{4,\delta}^{A}(R) - \cG_{4,\delta}^{A}(R)$ for $A \in \{\mathrm{anti-}k_t, k_t, \mathrm{C/A}\}$. Numerical values of $\Phi_{4,\delta}^{A}/4!$ for both the $k_t$ and C/A algorithms, together with the pure C/A correction $\Delta\Phi_{4,\delta}/4! \equiv (\Phi_{4,\delta}^{\caa} - \Phi_{4,\delta}^{\ktt})/4!$, are given in Table~\ref{tab:FG4-Values} at representative values of $R$.  All values are obtained by numerical integration using the   \texttt{Cuba} library.   A negative value of $\Phi_{4,\delta}^{A}/4!$ indicates that NGLs   dominate CLs ($\cG_{4,\delta}^{A} > \cF_{4,\delta}^{A}$); the   single positive C/A entry for channel~$(\delta_3)$ at $R=1.0$   signals a local reversal of this hierarchy.   A positive $\Delta\Phi_{4,\delta}/4!$ means C/A   reduces the non-global coefficient relative to $k_t$ and a negative value means the opposite.
\begin{table}[t]
\centering
\caption{Numerical values of the combined four-loop CL and NGL coefficient   $\Phi_{4,\delta}^{A}/4!$, Eqs.~\eqref{eq:4loop:CLs-NGLs} and~\eqref{eq:4loop:combined}, for the $k_t$ and C/A   algorithms, and the pure C/A correction   $\Delta\Phi_{4,\delta}/4!$,  at representative jet radii $R$ for all three Born channels. }
\label{tab:FG4-Values}
\setlength{\tabcolsep}{7pt}
\renewcommand{\arraystretch}{1.15}
\begin{tabular}{%
c c S[table-format=-1.2]   S[table-format=-1.2]   S[table-format=+1.2] }
\toprule
$\mathbf{R}$ & \textbf{Channel}  & {\bf $\Phi_{4,\delta}^{\ktt}/4!$}  & {\bf $\Phi_{4,\delta}^{\caa}/4!$} & {\bf $\Delta\Phi_{4,\delta}/4!$} \\
\midrule
\multirow{3}{*}{$0.4$}
  & $(\delta_1)$ & -2.37 & -1.82 & +0.55 \\
  & $(\delta_2)$       & -1.85 & -1.78 & +0.07 \\
  & $(\delta_3)$         & -4.78 & -6.50 & -1.72 \\
\midrule
\multirow{3}{*}{$0.6$}
  & $(\delta_1)$ & -2.45 & -1.93 & +0.52 \\
  & $(\delta_2)$ & -2.49 & -2.42 & +0.07 \\
  & $(\delta_3)$ & -3.96 & -5.70 & -1.74 \\
\midrule
\multirow{3}{*}{$0.7$}
  & $(\delta_1)$ & -2.09 & -1.46 & +0.63 \\
  & $(\delta_2)$ & -3.46 & -3.14 & +0.32 \\
  & $(\delta_3)$ & -4.32 & -5.75 & -1.43 \\
\midrule
\multirow{3}{*}{$0.8$}
  & $(\delta_1)$ & -2.22 & -1.63 & +0.59 \\
  & $(\delta_2)$ & -4.26 & -3.46 & +0.80 \\
  & $(\delta_3)$ & -4.07 & -4.62 & -0.55 \\
\midrule
\multirow{3}{*}{$1.0$}
  & $(\delta_1)$                       & -1.84 & -1.07 & +0.77 \\
  & $(\delta_2)$                       & -4.74 & -2.30 & +2.44 \\
  & $(\delta_3)$                      & -0.98 & +1.22 & +2.20 \\
\bottomrule
\end{tabular}
\end{table}

\begin{figure}[t]
\centering
\includegraphics[scale=0.55]{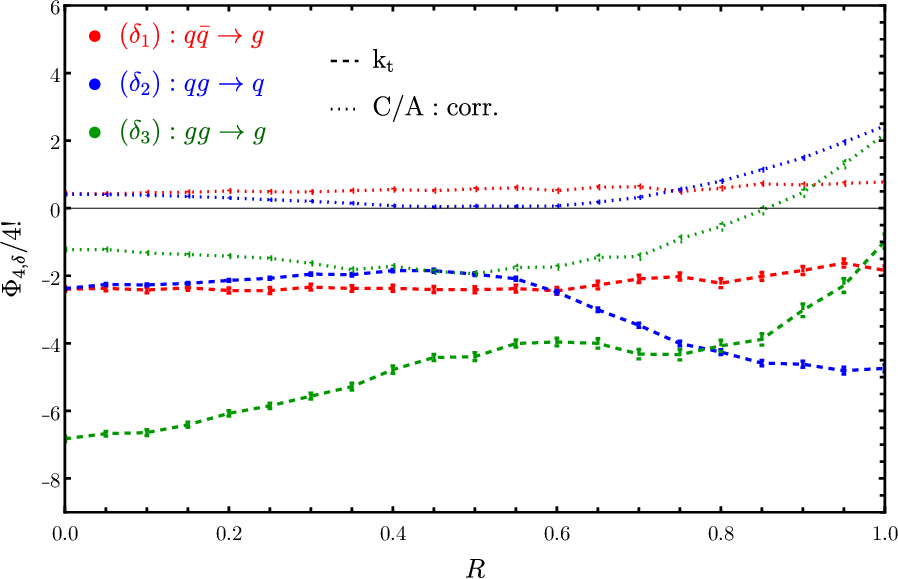}
\caption{The combined CL and NGL coefficient for the pure C/A corrections (dotted) and the full $k_t$ clustering (dashed) at four-loop order, shown as a function of $R$ for all three Born channels.}
\label{fig:GF4a}
\end{figure}

Fig.~\ref{fig:GF4a} shows the combined quantity $\Phi_{4, \delta}/4!$ for the pure C/A corrections alongside the $k_t$ results at four-loop order, for all three channels as a function of the jet radius $R$. The combined effect of CLs and NGLs for the pure C/A correction is relatively small and of opposite sign to the $k_t$ result for channels $(\delta_1)$ and $(\delta_2)$,  confirming that C/A consistently reduces the non-global logarithmic contribution relative to $k_t$ for these two channels. The corresponding result for the pure gluon-fusion channel $(\delta_3)$ is both relatively large (reaching $\Delta\Phi_{4,\delta_3}/4!
\approx -1.74$ at $R=0.6$; see Table~\ref{tab:FG4-Values}) and of the same sign as the $k_t$ result up to $R \sim 0.85$ where it vanishes and flips sign.

Fig.~\ref{fig:GF4c} depicts the combined CL and NGL effect at four-loop order, $\Phi_{4, \delta}$, for all three channels as a function of $R$, comparing all three sequential recombination algorithms: anti-$k_t$, $k_t$, and C/A. The following observations are worth noting:
\begin{figure}[t]
\centering
\includegraphics[scale=0.55]{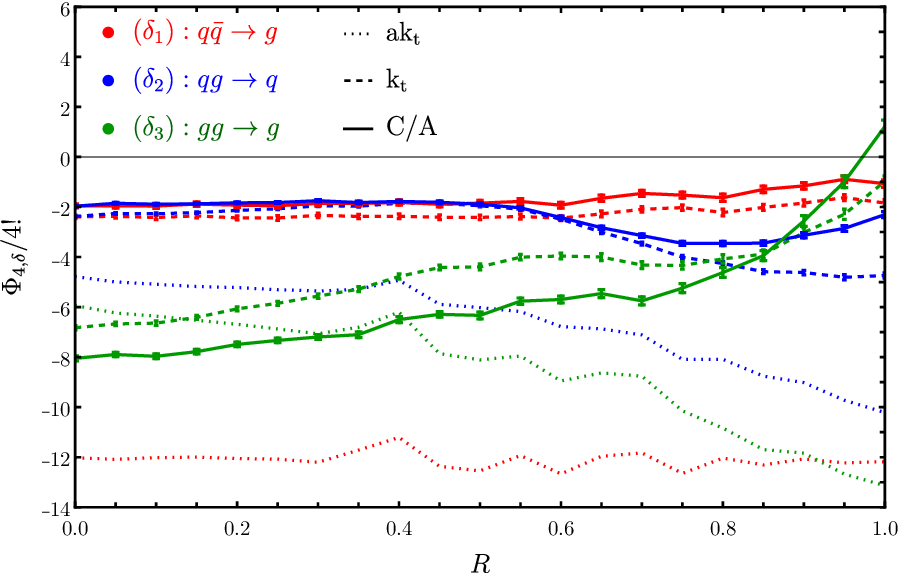}
\caption{The combined CL and NGL coefficient for the full C/A (solid), full $k_t$ (dashed), and anti-$k_t$ (dotted) clusterings at four-loop order, shown as a function of $R$ for all three Born channels.}
\label{fig:GF4c}
\end{figure}
\begin{itemize}
\item Channels $(\delta_2)$ and $(\delta_3)$ follow the pattern seen at three-loop order: both exhibit a reduced non-global logarithmic impact compared to both anti-$k_t$ and $k_t$, though no optimal jet radius exists at which this impact vanishes entirely. Channel $(\delta_2)$ for C/A is nonetheless comparable to $k_t$ over a wide range of $R$. The reduction in the combined CL and NGL coefficient reaches 84\% and 59\% for channels $(\delta_1)$ and $(\delta_2)$, respectively, compared to the anti-$k_t$ case, while the corresponding reductions for $k_t$ are approximately 80\% and 50\%. This confirms the relatively small corrections induced by C/A clustering compared to $k_t$. At the phenomenologically relevant jet radius $R = 0.7$, the reductions are 88\% and 56\% for C/A and 82\% and 51\% for $k_t$ for these two channels, further confirming the superiority of the C/A algorithm in suppressing non-global effects over a wide range of jet radii.

\item Channel $(\delta_3)$, on the other hand, deviates from the three-loop pattern. For $R \lesssim 0.4$, the combined non-global effect for C/A exceeds that for both anti-$k_t$ and $k_t$ by at most 35\% (at $R = 0$). For $R \gtrsim 0.4$, the non-global effect for C/A is continuously reduced relative to anti-$k_t$, reaching a reduction of approximately 100\% at $R \sim 0.97$, corresponding to the complete elimination of this effect; it then changes sign beyond this jet radius. Compared to $k_t$ clustering, the non-global effect for C/A is larger for $R \lesssim 0.9$. Hence, for channel $(\delta_3)$, the $k_t$ algorithm is preferred to C/A in suppressing the impact of non-global logarithms.

\item In the small-$R$ limit, channels $(\delta_1)$ and $(\delta_2)$ overlap at values of approximately $-1.96$ and $-2.38$ for C/A and $k_t$ clusterings, respectively. This differs from the two- and three-loop cases, in which channels $(\delta_1)$ and $(\delta_3)$ overlap in the small-$R$ limit. This indicates that the gluon-fusion channel behaves in an unexpected manner starting at four-loop order. Indeed, it has been observed in previous works (see, e.g.,~\cite{Delenda:2015tbo,Khelifa-Kerfa:2015mma}) that several patterns established at lower loop orders break down at four loops, both at the level of eikonal amplitudes and, consequently, at the cross-section level. It is also worth noting that the values obtained in the limit $R \to 0$ differ from those found in the $e^+e^-$ dijet case~\cite{Khelifa-Kerfa:2025jzl}.
\end{itemize}
An appealing feature of the three-loop results is that the combined CL and NGL coefficient vanishes at specific values of the jet radius
for channels~$(\delta_2)$ and~$(\delta_3)$ in C/A clustering ($R \approx 0.55$--$0.58$, Fig.~\ref{fig:GF3}), raising the prospect of an ``optimal'' radius at which non-global logarithms are eliminated. However, several issues temper this observation.  First, even at three-loop order, channel~$(\delta_1)$ admits no such zero; thus, no single value of~$R$ simultaneously suppresses non-global effects across all three Born channels.  The phenomenologically relevant quantity is the Born-cross-section-weighted combination
\begin{align}
\propto \sum_\delta\int\mathrm{d}\cB_\delta\, (\cF_{3,\delta}^{\caa} - \cG_{3,\delta}^{\caa}),
\end{align}
which is not computed here and is unlikely to vanish at the same~$R$ as the individual channels.  Second, and more fundamentally, the vanishing of the three-loop coefficient at $R\approx 0.55$ is an order-by-order cancellation between CLs and NGLs, not a signal of a physical suppression mechanism: the four-loop contributions to the same channels are non-zero at that radius (Fig.~\ref{fig:GF4c} and Table~\ref{tab:FG4-Values}), confirming that the cancellation does not persist beyond three-loop order.  Establishing whether a physically meaningful optimal radius exists at all orders would require either fixed-order calculations beyond four loops --- a task that is currently infeasible given the rapidly growing complexity of the eikonal amplitudes, the intricate geometry and combinatorics of multi-gluon configurations, and the sheer number of phase-space orderings that must be enumerated --- or a dedicated all-orders numerical resummation for C/A clustering. The latter path is not free of obstacles: the only available all-orders C/A code~\cite{Becher:2023znt} is restricted to purely soft observables (such as the energy-flow observable), since the BMS equation~\cite{Banfi:2002hw} on which it rests is formulated under the assumption of purely soft singularities.  The invariant jet-mass observable exhibits both soft and collinear singularities, which places it outside the scope of both the BMS framework and the aforementioned  numerical code. The Dasgupta--Salam code seems, currently, the only promising short-term avenue and might be modified to cover the C/A clustering.  Such a task will be pursued in future publications. Another, long-term option is to extend the BMS equation to cases where the final-state soft partons can be clustered with one of the sequential recombination jet algorithms.

To quantify the phenomenological impact of C/A clustering, it is instructive to perform an all-orders estimate of the effect of the computed non-global logarithms on the resummed invariant jet-mass distribution. This is the subject of the next section.

%%%%%%%%%%%%%%%%%%%%%%%%%%%%%%%%%%%%%%%%%%%%%%%%%%%%%%%%%%
\section{Estimation of all-orders effects}
\label{sec:AllOrders}

In order to estimate the contribution of the missing higher-order terms in the CL and NGL series to the invariant jet-mass distribution, and to investigate the impact of the non-global part on the overall distribution, we follow the standard procedure (see, for instance,~\cite{Catani:1992ua, Banfi:2010pa, Khelifa-Kerfa:2011quw, Kerfa:2012yae, Khelifa-Kerfa:2024udm, Khelifa-Kerfa:2025jev, Khelifa-Kerfa:2025jzl}). Specifically, based on the fixed-order perturbative calculations up to four-loop order presented above, we assume that both CLs and NGLs exponentiate, and that their resummed form factors appearing in Eq.~\eqref{eq:fB_decomp} take, at NLL accuracy, the form
\begin{subequations}
\label{eq:nloop:S-C-FormFactors}
\begin{align}
\cC_{\delta}^{\caa}(t) &= \exp\!\left[\sum_{n=2}^4 \frac{(-1)^n}{n!}\,\cF_{n,\delta}^{\caa}(R)\,(2t)^n\right], \\
\cS_{\delta}^{\caa}(t) &= \exp\!\left[-\sum_{n=2}^4 \frac{(-1)^n}{n!}\,\cG_{n,\delta}^{\caa}(R)\,(2t)^n\right],
\end{align}
\end{subequations}
where $\cF_{n,\delta}^{\caa}$ and $\cG_{n,\delta}^{\caa}$ are the CL and NGL coefficients, respectively, computed in the previous section. The evolution variable $t$ is related to the invariant jet-mass observable $\rho$ at one-loop order --- sufficient for NLL accuracy --- by
\begin{figure*}[t]
\centering
\includegraphics[scale=0.55]{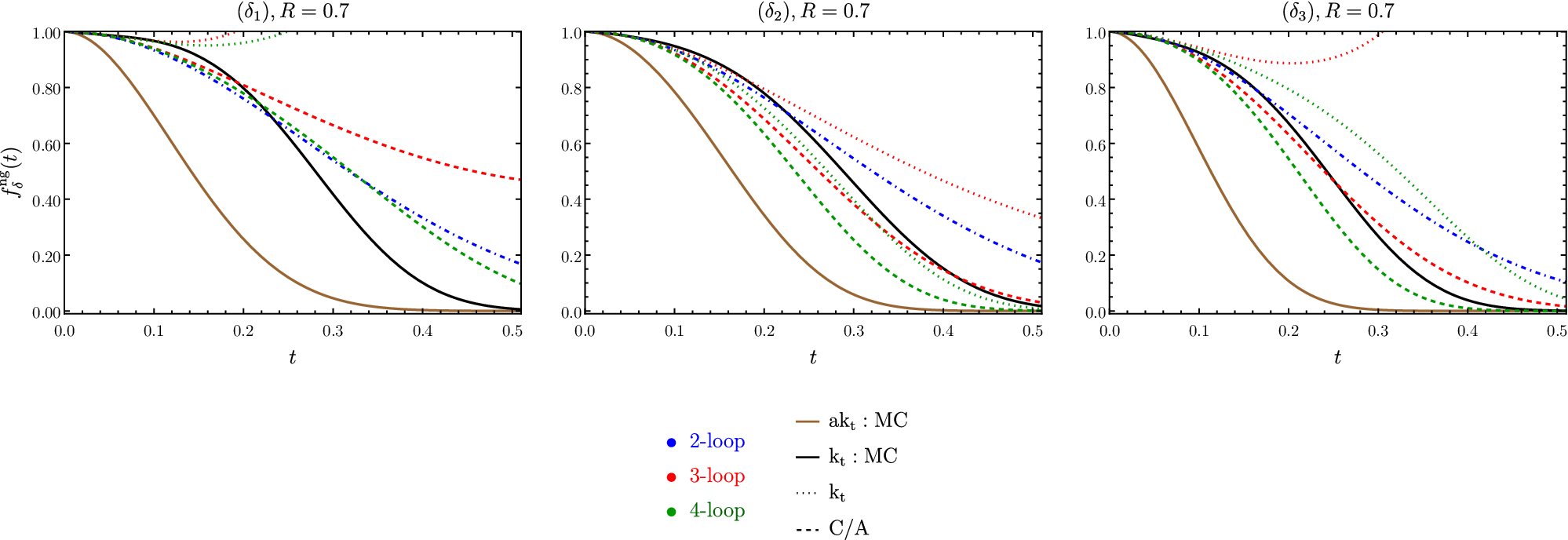}
\caption{Comparisons of the fixed-order exponentiated results at two-, three-, and four-loop order for the $k_t$ (dotted) and C/A (dashed) algorithms against the output of the MC code of~\cite{Dasgupta:2001sh} for the anti-$k_t$ (brown solid) and $k_t$ (black solid) algorithms. Results are shown for jet radius $R = 0.7$ and for all three Born channels.}
\label{fig:fR7}
\end{figure*}
\begin{equation}
t = -\frac{1}{4\pi\beta_0}\ln\!\left[1 - 2\asb(p_t)\beta_0 L\right],
\label{eq:t-evol}
\end{equation}
with $L = \ln(R^2/\rho)$ and the leading-order QCD beta-function coefficient $\beta_0 = (11\CA - 2n_f)/(12\pi)$, where $n_f$ is the number of active quark flavors. These fixed-order-based resummed form factors are then compared to the output of specialized numerical Monte Carlo codes, such as the one developed by Dasgupta and Salam~\cite{Dasgupta:2001sh}. The latter are parametrized as follows:
\begin{subequations}
\label{eq:MC-FormFactors}
\begin{align}
\cC_{\delta}^{\MC}(t) &= \exp\!\left[\sum_{(ij)\in\Delta_\delta} \cC_{ij}^2\,\cF_{2,\delta}^{(ij)}(R)\,f_{ij}(t)\right], \\
\cS_{\delta}^{\MC}(t) &= \exp\!\left[-\CA\sum_{(ij)\in\Delta_\delta} \cC_{ij}\,\cG_{2,\delta}^{(ij)}(R)\,\tilde{f}_{ij}(t)\right],
\end{align}
\end{subequations}
where $\cF_{2,\delta}^{(ij)}$ and $\cG_{2,\delta}^{(ij)}$ are the two-loop CL and NGL coefficients, respectively, computed in our previous paper~\cite{Khelifa-Kerfa:2025jev} (recall that the C/A coefficients are identical to the $k_t$ ones at this loop order). The Born color factors $\cC_{ij}$ are defined in Eq.~\eqref{eq:dipolecolor}. The fitting functionals $f_{ij}$ and $\tilde{f}_{ij}$ are given by
\begin{equation}
f_{ij}(t) = \frac{1 + (\lambda_{ij}\,t)^2}{1 + (\sigma_{ij}\,t)^{\gamma_{ij}}}\,t^2.
\label{eq:fitting-func}
\end{equation}
Both $f_{ij}$ and $\tilde{f}_{ij}$ share the same functional form but differ in the values of the fitting parameters $\lambda_{ij}$, $\sigma_{ij}$, and $\gamma_{ij}$. The values of these parameters for the $k_t$ case at $R = 0.7$ are given in Eqs.~(4.3) and~(4.6) of Ref.~\cite{Khelifa-Kerfa:2025jev}.

As noted previously, the MC code of~\cite{Dasgupta:2001sh} does not implement the C/A jet algorithm. Moreover, the numerical code of~\cite{Becher:2023znt}, which does implement all three sequential recombination algorithms including C/A, is restricted to observables with purely soft singularities and does not cover observables with simultaneous soft and collinear singularities, such as the invariant jet-mass studied here. Both codes furthermore operate in the large-$N_c$ limit, whereas our analytical calculations retain the full finite-$N_c$ color structure; no all-orders numerical tool covering C/A clustering at finite $N_c$ for this class of observables is currently available. The only existing all-orders numerical result at finite $N_c$ concerns the energy-flow observable in Higgs decays for the anti-$k_t$ algorithm~\cite{Hatta:2020wre}. We therefore compare our analytical calculations to the output of the MC code of~\cite{Dasgupta:2001sh} for the anti-$k_t$ and $k_t$ algorithms, which serves to estimate the impact of C/A clustering at all orders relative to those two algorithms. A similar analysis was performed in the $e^+e^-$ dijet process in our previous work~\cite{Khelifa-Kerfa:2025jzl}.

Since at four-loop order the CLs and NGLs could not be decoupled from each other, we plot, in Fig.~\ref{fig:fR7}, the products $f_{\delta}^{\mathrm{ng}}(t) = \cC_{\delta}(t)\times\cS_{\delta}(t)$ and $\cC_{\delta}^{\MC}(t)\times\cS_{\delta}^{\MC}(t)$ for all three jet algorithms and all Born channels. Recall that for the anti-$k_t$ algorithm there is no CL contribution, so $\cC_{\delta}^{\aktt}(t) = 1$. The labels ``2-loop'', ``3-loop'', etc.\ correspond to truncating the series in the exponents of Eqs.~\eqref{eq:nloop:S-C-FormFactors} at $n = 2$, $3$, etc.

Recalling that $f_{\delta}^{\mathrm{ng}}$ multiplies the Sudakov form factor in Eq.~\eqref{eq:fB_decomp}, Fig.~\ref{fig:fR7} clearly shows, as anticipated, that the anti-$k_t$ algorithm produces the most severe suppression of the total distribution --- that is, the largest reduction in the overall size of the invariant jet-mass distribution across all three channels. The $k_t$ algorithm produces a considerably smaller reduction over a wide range of the variable $t$, making it a more favorable choice as it more effectively mitigates the net impact of non-global logarithms and brings the total distribution closer to the pure Sudakov result (see Ref.~\cite{Khelifa-Kerfa:2025jev} for further details).
\begin{figure*}[t]
\centering
\includegraphics[scale=0.7]{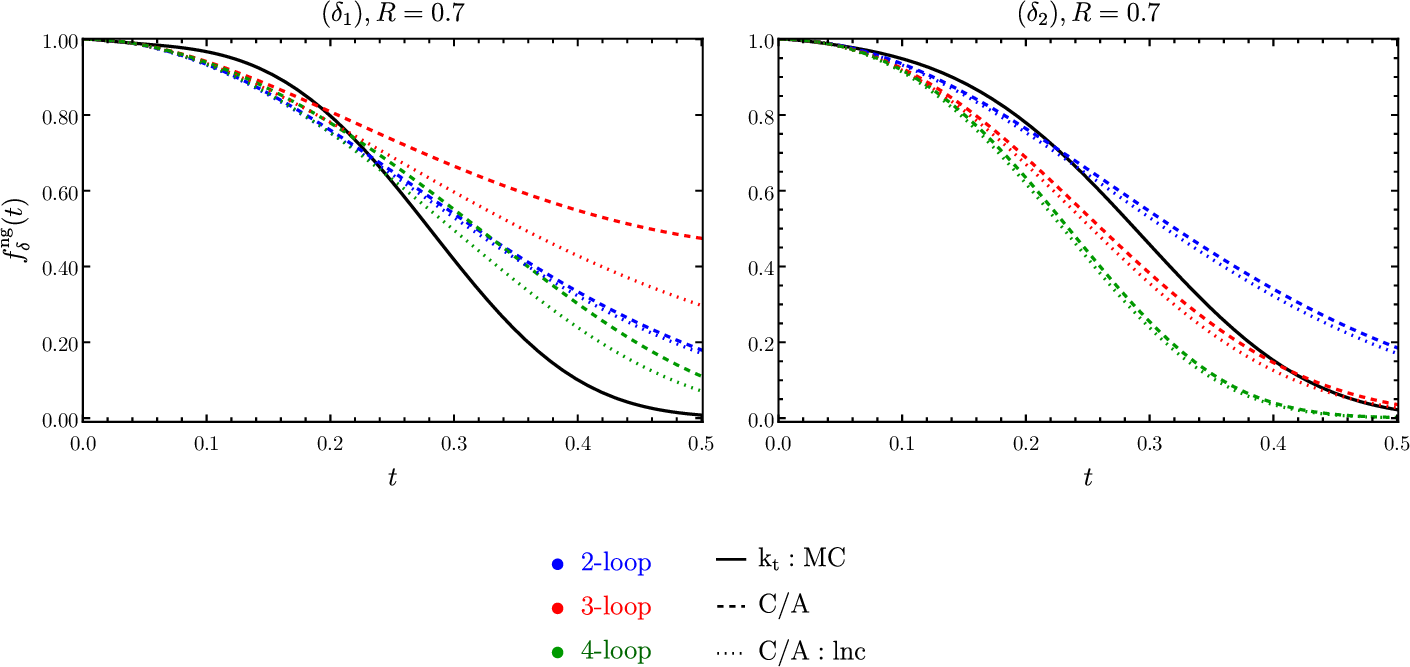}
\caption{Comparisons of the fixed-order exponentiated results at two-, three-, and four-loop order for the C/A jet algorithm at finite- (dashed) and large-$N_c$ (dotted).  Results are shown for jet radius $R = 0.7$ and for the two distinct Born channels.}
\label{fig:fR7-nc}
\end{figure*}
The exponentiation of the fixed-order results is reliable for small values of $t$, specifically $t \lesssim 0.15$, corresponding to $\rho \gtrsim 0.005$. This is confirmed by comparing the $k_t$ fixed-order exponentiation against the $k_t$ MC curve, for which both are available. In this range, the C/A effect is comparable to the $k_t$ one. For larger values of $t$, where the fixed-order exponentiation is no longer reliable, the C/A effect appears more aggressive than $k_t$, shifting the curves toward the anti-$k_t$ result. This apparent contradiction with the three- and four-loop fixed-order findings --- which show C/A outperforming $k_t$ in suppressing non-global logarithms --- further confirms that Fig.~\ref{fig:fR7} should only be trusted in the range $t \lesssim 0.15$--$0.2$, which is the phenomenologically relevant region. As noted in previous works (see, for instance,~\cite{Banfi:2010pa,Dasgupta:2012hg,Khelifa-Kerfa:2015mma}), the two-loop results provide the best approximation to the all-orders MC curves over a wide range of $t$. Given the absence of an all-orders numerical estimate for C/A clustering, our analysis shows that the C/A and $k_t$ algorithms perform almost equally in reducing the impact of non-global logarithms at all orders. Nonetheless, the fixed-order calculations clearly indicate a preference for C/A over $k_t$ in this regard.

Fig.~\ref{fig:fR7-nc} quantifies the effect of finite-$N_c$ corrections on the exponential NGL form factor constructed from the fixed-order coefficients, for channels $(\delta_1)$ and $(\delta_2)$ at $R = 0.7$. Recall that the third all--gluon Born channel $(\delta_3)$ receives no finite-$N_c$ corrections. For the phenomenologically relevant range $t \leq 0.15$, the finite-$N_c$ corrections remain below 2\% for both channels, in agreement with previous findings~\cite{Hatta:2013iba,Hatta:2020wre,Khelifa-Kerfa:2024udm}.

\begin{figure}[t]
\centering
\includegraphics[scale=0.56]{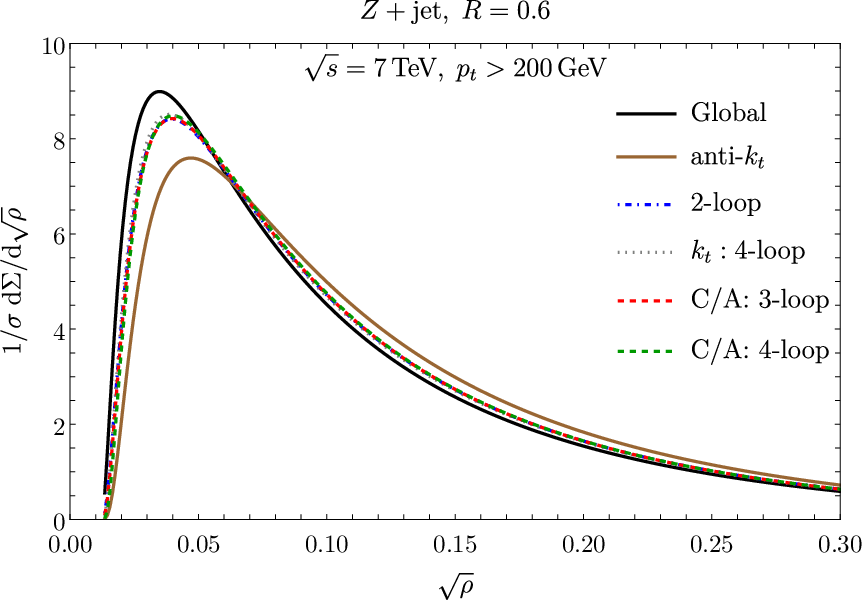}
\caption{Various approximations to the normalized differential resummed distribution in the (square-root of the) invariant jet-mass squared for the leading-$p_t$ jet in $Z$+jet production at $\sqrt{s} = 7\,\text{TeV}$ and $R = 0.6$. The $k_t$ curves are shown for reference. The two-loop exponential is identical for both $k_t$ and C/A; the three- and four-loop results differ between the two algorithms.}
\label{fig:ZjR6}
\end{figure}

To assess the phenomenological impact of non-global logarithms in the C/A clustering more concretely, we show in Fig.~\ref{fig:ZjR6} the normalized invariant jet-mass differential cross section of the leading-$p_t$ ($p_t > 200\,\text{GeV}$) jet, summed over all three Born channels, for the specific hadronic process $Z$+jet at $\sqrt{s} = 7\,\text{TeV}$ and $R = 0.6$.
As anticipated from Fig.~\ref{fig:fR7}, the impact of C/A clustering is of the same magnitude as that of $k_t$ clustering. The position of the peak is shifted slightly to the right by approximately the same amount at all three loop orders, with the reduction in peak height decreasing from approximately 6.43\% at two-loop order to 5.67\% at four-loop order. These values are nearly identical to those found for the $k_t$ case in Ref.~\cite{Khelifa-Kerfa:2025jev}. Compared to the anti-$k_t$ clustering, which induces a reduction of approximately 15.5\% in the peak height, the C/A and $k_t$ algorithms perform comparably and both significantly better at reducing the overall impact of non-global logarithms.

%%%%%%%%%%%%%%%%%%%%%%%%%%%%%%%%%%%%%%%%%%%%%%%%%%%%%%%%%%
\section{Conclusion}
\label{sec:Conclusion}

The present paper is an extension of our previous series of works on the structure and impact of non-global logarithms appearing in the distribution of generic non-global jet observables, where the final-state jets are defined using the three well-known sequential recombination algorithms: anti-$k_t$, $k_t$, and C/A. Those calculations covered jet production in both leptonic and hadronic environments and included the full jet-radius dependence as well as finite-$N_c$ corrections. They were carried out at fixed perturbative order up to four loops. The observable distribution clearly exhibits an exponentiation pattern, which is exploited to construct an all-orders estimate in terms of the coefficients computed at each loop order. The resulting resummed distributions are then validated against the output of numerical MC codes wherever available.

In the present work we have computed the normalized invariant jet-mass distribution for the leading-$p_t$ jet in Higgs- and vector-boson production in association with a jet --- a process that has been widely studied by both the phenomenology and experimental communities. The final-state jets are clustered using the Cambridge/Aachen algorithm for various jet radii. Fixed-order calculations are performed through four-loop order in perturbation theory. Applying C/A clustering in this hadronic setting is substantially more complex than the $k_t$ case, owing to the fact that the C/A distance measure is independent of the transverse momenta of the final-state partons; consequently, all possible orderings of the pairwise distances must be taken into account. This task has been fully automated using scripts written in \texttt{Python} together with \texttt{Mathematica}.

At two-loop order, the C/A clustering is equivalent to $k_t$ and the results are therefore identical to those reported in our previous work~\cite{Khelifa-Kerfa:2025jev}. At three- and four-loop order, the C/A algorithm produces contributions that differ from those of $k_t$, and are presented here for the first time in literature. In general, the effect of non-global logarithms on the total observable distribution is smaller for C/A than for $k_t$, a feature that holds for CLs, NGLs, and their combined effect. The four-loop calculation is the most demanding, and as a result a separation of the CL and NGL contributions has not been possible at this order. Nevertheless, it is their combined effect that ultimately enters the observable distribution.

The all-orders analysis reveals that the resummed form factor constructed from the fixed-order calculations can be reliably trusted only for $\rho \gtrsim 0.005$ (corresponding to $m_j^2 \gtrsim 0.005 \,p_t^2$). Furthermore, the overall impact of non-global logarithms for C/A clustering has been shown to be of the same order as that for $k_t$ clustering. Both algorithms perform significantly better than anti-$k_t$ in suppressing the impact of these logarithms on the final observable distribution over a wide range of jet radii.

The present work completes the fixed-order study of the effect of the three sequential recombination algorithms on non-global observables through four-loop order, for the specific processes involving two and three hard colored legs, namely $e^+e^-$ annihilation and $V/H$+jet production. Future work should address higher loop orders and/or tackle the all-orders resummation directly, since it appears difficult to reliably infer the latter from fixed-order calculations alone. Other hadronic processes that warrant careful investigation include extending the multi-loop analysis to dijet production, which involves a considerably more complex color and kinematic structure.
%%%%%%%%%%%%%%%%%%%%%%%%%%%%%%%%%%%%%%%%%%%%%%%%%%%%%%%
%\begin{acknowledgments}
%
%\end{acknowledgments}

%%%%%%%%%%%%%%%%%%%%%%%%%%%%%%%%%%%%%%%%%%%%%%%%%%%%%%%
%%\appendix
%
%%%%%%%%%%%%%%%%%%%%%%%%%%%%%%%%%%%%%%%%%%%%%%%%%%%%%%%
%Bib
\bibliography{Refs}

\end{document}